\documentclass[useAMS,twocolumn,usenatbib]{mn2e}

\usepackage{graphicx}
\usepackage{amssymb,amsmath}
\usepackage{aas_macros}

\usepackage[normalem]{ulem}
\usepackage[usenames]{color}

\begin{document}

\title[The Lyman-$\alpha$ forest in a blazar-heated Universe]
{The Lyman-$\alpha$ forest in a blazar-heated Universe}
\author[E. Puchwein, C. Pfrommer, V. Springel, A. E. Broderick \& P. Chang]
{Ewald Puchwein$^1$\thanks{E-mail: ewald.puchwein@h-its.org},
Christoph Pfrommer$^1$,
Volker Springel$^{1,2}$,\newauthor
Avery E. Broderick$^{3,4,5}$,
and Philip Chang\vspace{0.15cm}$^{3,6}$
\\$^1$Heidelberg Institute for Theoretical Studies, Schloss-Wolfsbrunnenweg 35, D-69118 Heidelberg, Germany
\\$^2$Zentrum f\"ur Astronomie der Universit\"at Heidelberg, Astronomisches
Recheninstitut, M\"{o}nchhofstr. 12-14, 69120 Heidelberg, Germany
\\$^3$Canadian Institute for Theoretical Astrophysics, 60 St.~George Street, Toronto, ON M5S 3H8, Canada
\\$^4$Perimeter Institute for Theoretical Physics, 31 Caroline Street North, Waterloo, ON, N2L 2Y5, Canada
\\$^5$Department of Physics and Astronomy, University of Waterloo, 200 University Avenue West, Waterloo, ON, N2L 3G1, Canada
\\$^6$Department of Physics, University of Wisconsin-Milwaukee, 1900 E. Kenwood Boulevard, Milwaukee, WI 53211, USA
}
\date{\today}
\maketitle

\begin{abstract} 
It has been realised only recently that TeV emission from blazars can significantly heat the intergalactic medium (IGM) by pair-producing high-energy electrons and positrons, which in turn excite vigorous plasma instabilities, leading to a local dissipation of the pairs' kinetic energy. In this work, we use cosmological hydrodynamical simulations to model the impact of this blazar heating on the Lyman-$\alpha$ forest at intermediate redshifts ($z\sim2-3$). We find that blazar heating produces an inverted temperature-density relation in the IGM and naturally resolves many of the problems present in previous simulations of the forest that included photoionisation heating alone. In particular, our simulations with blazar heating simultaneously reproduce the observed effective optical depth and temperature as a function of redshift, the observed probability distribution functions (PDFs) of the transmitted flux, and the observed flux power spectra, over the full redshift range $2<z<3$ analysed here. Additionally, by deblending the absorption features of Lyman-$\alpha$ spectra into a sum of thermally broadened individual lines, we find superb agreement with the observed lower cutoff of the line-width distribution and abundances of neutral hydrogen column densities per unit redshift. Using the most recent constraints on the cosmic ultraviolet (UV) background, this excellent agreement with observations does not require rescaling the amplitude of the UV background; a procedure that was routinely used in the past to match the observed level of transmitted flux. We also show that our blazar-heated model matches the data better than standard simulations even when such a rescaling is allowed. This concordance between Lyman-$\alpha$ data and simulation results, which are based on the most recent cosmological parameters, also suggests that the inclusion of blazar heating alleviates previous tensions on constraints for $\sigma_8$ derived from Lyman-$\alpha$ measurements and other cosmological data. Finally, we show that blazar heating dramatically alters the volume-weighted temperature PDF, implying an important change of the strengths of structure formation shocks (and thereby possibly particle acceleration in these shocks). The density PDF is also modified, suggesting that blazar heating may have interesting effects on structure formation, particularly on the smallest galaxies. 
\end{abstract}


\section{Introduction}
\label{sec:introduction}

The thermal history of the intergalactic medium (IGM) is of key importance for understanding cosmological structure formation since it determines the initial thermodynamic conditions for the formation of collapsed objects ranging from dwarf galaxies up to the scale of galaxy clusters. Simultaneously, the IGM acts as a cosmic heat reservoir that records reionisation processes in the Universe that inject substantial amounts of energy into the IGM on comparably short cosmological time scales. Detailed measurements of the thermal history of the IGM may therefore reveal the main contributors to the UV background, e.g., cosmic star formation or quasars, which are thought to be primarily responsible for these reionisation events. 

Until recently, it was generally thought that all the main sources for heating the IGM had been identified and only the detailed interplay of these processes remained to be understood.  However, in a recent series of papers \citep{Broderick2011,Chang2011,Pfrommer2011}, a novel heating process was proposed that taps efficiently into the energy reservoir of TeV blazars which are ultimately powered by accretion onto their central super-massive black holes (or more generally the engines of active galactic nuclei, AGN).  A subset of blazars emits the bulk of their energy in TeV $\gamma$-ray radiation ($E\gtrsim 100~\rmn{GeV}$) to which the Universe is opaque; i.e., these energetic gamma rays necessarily annihilate upon the photons comprising the extragalactic background light (EBL), producing an ultrarelativistic population of electron-positron pairs. Typical gamma-ray mean free paths range from 30 Mpc to 1 Gpc, depending upon $\gamma$-ray energy and source redshift. The pairs produced by TeV $\gamma$-rays have typical Lorentz factors of $10^5-10^7$. Despite its dilute nature, a beam of these ultra-relativistic pairs is susceptible to plasma beam instabilities while propagating through the IGM. In particular, the ``oblique'' instability, a more virulent and robust cousin of the commonly discussed Weibel and two-stream instabilities, appears to grow out of the pair beam anisotropy as a source of free energy \citep{Bret-Firp-Deut:04,Bret-Firp-Deut:05,Bret:09,Bret-Grem-Diec:10,Lemo-Pell:10}. Though the nonlinear evolution of these instabilities is unknown, they may saturate at a level such that they dissipate the kinetic energy of the beam at a rate comparable to their linear growth rate, thereby heating the IGM locally, following the suggestion of \citet{Broderick2011}.

If this scenario or an analogous, similarly efficient mechanism operates in practise, it necessarily suppresses the inverse Compton cooling of the pairs, which acts on a longer timescale. This naturally explains the absence of an inverse Compton bump in observed TeV-blazar spectra without invoking an intergalactic magnetic field \citep[][and references therein]{Broderick2011}. At the same time, it allows for a redshift evolution of TeV blazars that is identical to that of quasars without overproducing the extragalactic $\gamma$-ray background (EGRB).  In fact, for plausible parameters of TeV-blazar spectra, it is possible to explain the high-energy part of the EGRB \citep{Broderick2011}.  This is in stark contrast to the previous picture which assumed that the TeV $\gamma$-ray emission is reprocessed into the GeV regime by means of inverse Compton processes. In this picture, in order to not overproduce the EGRB observed by {\em Fermi} \citep{Fermi_EGRBApJ2010}, the spectral redistribution of TeV $\gamma$-ray energy into the GeV regime strongly constrains the evolution of the luminosity density of TeV blazars. In these studies it was found that TeV blazars cannot exhibit the dramatic rise in number seen in the quasar distribution by $z\sim1$--$2$, i.e., the comoving number of blazars must have remained essentially fixed \citep[see, e.g.,][]{Naru-Tota:06,Knei-Mann:08,Inou-Tota:09,Vent:10}; at odds with both the large-scale mass assembly history of the Universe as traced by the star formation history, and with the luminosity history of similarly accreting systems, e.g., the quasar luminosity density.

Integrating over the energy flux per mean free path of all known TeV blazars yields a luminosity density, or equivalently a local heating rate, that dominates that of photoheating by more than an order of magnitude at the present epoch at mean density, after accounting for incompleteness corrections \citep{Chang2011}. As demonstrated in \citet{Broderick2011}, the local TeV blazar luminosity function is consistent with a scaled version of the quasar luminosity function \citep{Hopkins+07}, thus the conservative assumption is that they evolve similarly, presumably due to the same underlying accretion physics.  With this assumption, blazar heating starts to become important around the epoch of He {\sc ii} reionisation ($z\sim3.5$). By $z=0$, the blazar heating rate is more than an order of magnitude larger in comparison with the photoheating rate alone. 

Because the Universe is opaque to the propagation of TeV photons over cosmological distances, the redshift evolution of the TeV blazar luminosity density is observationally inaccessible for $z\gtrsim 0.5$. Also, the exact value of the blazar heating rate is subject to an uncertain incompleteness correction factor, which accounts for the incomplete sky coverage of current TeV instruments, the duty cycle of TeV blazars, their spectral variability, variations of the TeV-blazar redshift evolution, and additional TeV sources that may contribute to the plasma instability heating. Hence we can try to invert the problem and investigate how we can use the IGM as a cosmic calorimeter to probe the physics of TeV $\gamma$-rays and the global properties of TeV blazars by means of the energy that is deposited through plasma instabilities (or similarly efficient dissipation processes of the TeV emission from blazars).  In this manner, the thermal history of the IGM provides an analogous argument to that of \citet{Solt:82}, yielding an important constraint on the redshift evolution of the blazar luminosity density \citep{Chang2011}. Hence we turn to precision measurements of the Lyman-$\alpha$ forest at $z\sim 2-3$ \citep[e.g.,][]{Viel2009} in this work, with the goal to clarify the possible signatures of such an additional heating process, which is characterised by some unusual properties as we will discuss now.
 
Since the number densities of EBL photons and of TeV blazars are nearly homogeneous on cosmological scales, so is the resulting pair density. Hence, the implied heating rate is also homogeneous, i.e., the volumetric blazar heating rate is expected to be spatially uniform and {\em independent} of IGM density. This implies that the energy deposited per baryon is substantially larger in more tenuous parts of the Universe, such that underdense regions experience a larger temperature increase, producing an inverted temperature-density (T-$\rho$) relation with an asymptotic scaling of $T\propto 1/\rho$ for densities much lower than the cosmic mean. Interestingly, detailed studies of high-resolution Lyman-$\alpha$ forest spectra led to the conjecture that such an inverted T-$\rho$ relation in fact exists at $z=2-3$ \citep{Bolton2008,Viel2009}, something which has proven hard to arrange within the context of standard reionisation models \citep{McQuinn2009,Bolton2009}. Related to that, when deblending the absorption features of Lyman-$\alpha$ spectra into a sum of thermally broadened individual lines and measuring their strengths and widths, simulations produce line widths that are significantly narrower than observed, implying an associated inconsistency between the simulated and measured line width distribution \citep{Bryan1999,Bryan2000,Theuns1999,Machacek2000}. This finding is largely independent of the amount of small-scale power, but does depend strongly on the IGM temperature; heuristically increasing the temperature of the IGM can hence reproduce the observed line-width distribution rather well \citep{Theuns2000}.

In this work, we perform state-of-the-art simulations of the Lyman-$\alpha$ forest that include different variants of blazar heating (to address the uncertainty of the overall heating rate), and we compare the predictions to data in observational space. In this way, we obtain important information on the validity of the blazar heating scenario and we can elucidate whether it helps to resolve current problems in understanding some aspects of the IGM observations.  After describing our methods for simulating the Lyman-$\alpha$ forest in Sect.~\ref{sec:methods}, we present our results in Sect.~\ref{sec:results}. We focus on the thermal history and the T-$\rho$ relation of the IGM in Sect.~\ref{sec:th_eos_igm}, the effective optical depths and photoionisation rates in Sect.~\ref{sec:optical_depths}, the transmitted flux probability distribution functions and power spectra in Sect.~\ref{sec:flux_pdfs_power_spec}, and the Voigt profile fitting of the Lyman-$\alpha$ forest in Sect.~\ref{sec:line_fitting}. Finally, we give an outlook on implications for cosmological parameter estimation when including the physics presented in this work in Sect.~\ref{sec:outlook}, and conclude in Sect.~\ref{sec:conclusions}.

\section{Methods}
\label{sec:methods}

\subsection{The cosmological simulations}
\label{sec:simulations}

Throughout this paper, we make use of a set of cosmological hydrodynamical simulations that were performed with the TreePM-SPH simulation code {\sc P-Gadget-3}, which is an updated and extended version of {\sc Gadget-2} \citep{Springel2005b}. The simulations adopt the WMAP 7-year cosmology \citep{Komatsu2011} with $\Omega_{\rm M}=0.272$, $\Omega_{\Lambda} = 0.728$, $\Omega_{\rm B}=0.0465$, $h=0.704$, and $\sigma_{8}=0.809$. Initial conditions with $2 \times 128^3$ and $2 \times 384^3$ particles (half of them gas, the other half dark matter) in a cosmological box with a co-moving side length of $15  \, h^{-1} \, {\rm Mpc}$ were evolved from redshift $z_{\rm start}=100$ to redshift $z_{\rm end}=0$ or $z_{\rm end}=1.9$, respectively. 

Our analysis of the simulated Lyman-$\alpha$ forest is solely based on the higher resolution simulations with a fixed co-moving gravitational softening of $\epsilon = 1.5 \, h^{-1} \, {\rm kpc}$ (Plummer-equivalent) and dark matter and gas particle masses of $M_{\rm dm}=3.7\times10^6 \, h^{-1} M_\odot$ and $M_{\rm gas}= 7.5 \times 10^5 \, h^{-1} M_\odot$ respectively, while the lower resolution runs with a three times larger softening and 27 times larger particle masses were only used to study the overall thermal history of the universe.\footnote{Results from the $2 \times 128^3$ runs are only shown in Figs. \ref{fig:temp_evo} and  \ref{fig:T_rho_pdfs}. In the $2 \times 384^3$ simulations, the Lyman-$\alpha$ forest properties are well converged \citep[see e.g.,][for a discussion of the convergence of power spectra of the transmitted flux]{Viel2004}.} 

All of our runs follow hydrodynamics with an entropy-conserving formulation of smoothed-particle hydrodynamics (SPH) \citep[see][]{Springel2002}, and account for radiative cooling, star formation, and supernovae feedback with a sub-resolution model
described in \citet{Springel2003}. Photoheating was included under the assumption of ionisation equilibrium in the presence of an external UV background field as in \citet{Katz1996}, however, using a state-of-the-art model for the time evolution of the cosmic UV background \citep[based on][hereafter FG'09, corresponding to their Table 2]{Faucher-Giguere2009}. This UV background model agrees well with observational constraints from \citet{Faucher-Giguere2008} (hereafter FG'08).

Typically, when photoheating is followed in such a framework, hydrogen (H) and helium {\sc i} (He {\sc i}) reionisation happen very rapidly, which is problematic for two reasons: first, the assumption of ionisation equilibrium may not be satisfied during the reionisation transition itself, and second, the time steps in the simulation (which are limited by the Courant condition) may be too large to yield the correct overall temperature increase even if the equilibrium assumption were still valid, as the sharp peak in the heating rate may not be sampled well. This usually results in a too inefficient heating. We circumvent both problems by injecting the overall heat input expected for a sudden H and He {\sc i} reionisation by hand. For this, we need to assume values for the average excess energy per H and He {\sc i} ionisation event. We chose values that are consistent with the FG'09 model at $z=10$, namely, $E_{\rm H}=5.8 \, {\rm eV}$ and $E_{\rm He I}=9.6 \, {\rm eV}$. This results in a temperature increase of $\Delta T_{\rm H,He I} = 2.35\times 10^4 \, {\rm K}$.

One possible way in which to apply such a temperature boost would be to turn off the standard photoheating that assumes ionisation equilibrium with the UV background until $z\simeq 10$. We could then apply the temperature boost, $\Delta T_{\rm H,He I}$, to the SPH particles and subsequently turn on the standard photoheating to follow He {\sc ii} reionisation. At that time, our assumed UV background would be strong enough to keep H and He {\sc i} ionised. 

However, we adopted a different approach that allows us to bracket the uncertainties involved in the treatment of H and He {\sc i} reionisation. This involved performing two simulation runs where we did not disable the standard photoheating. In one of them we applied the temperature boost shortly after the reionisation by the UV background. This procedure somewhat overpredicts the heating since the standard photoheating implementation, despite being inefficient, will also have injected some energy. In the other run, we injected the energy slightly before the UV background reionises H and He {\sc i}. In this case, even after the injection, a fraction of the H and He will still be considered neutral given the assumption of ionisation equilibrium with the UV background. Thus, atomic cooling will remove part of the injected energy from the IGM until the UV flux has increased sufficiently to fully ionise H and He {\sc i}. Hence, in the second approach, we underestimate the temperature after H and He {\sc i} reionisation. Together, the two runs bracket the actual heat input. The thermal history of the IGM in these runs is discussed in more detail in Sect. \ref{sec:th_eos_igm}.

Finally, we would like to point out that He {\sc ii} reionisation happens much more gradually in the FG'09 model, with the bulk of the He {\sc ii} being reionised between redshifts 5 and 3. Hence, we are confident that the standard photoheating algorithm accurately follows the He {\sc ii} reionisation and yields the correct heating rates for the FG'09 model.\footnote{There is, however, still some uncertainty in the He {\sc ii} heating rate due to assuming optical-thin photoheating.}   

\subsection{Blazar heating}
\label{sec:blazar_model}

The main focus of this work is an investigation of the impact of blazar heating on the properties of the IGM and its observational signatures, particularly the Lyman-$\alpha$ forest. Thus, we performed four simulations at both resolutions, one without blazar heating and three with blazar heating at different levels of efficiency. All four runs were started from identical initial conditions, so they are directly comparable, allowing us to isolate and cleanly study the effects of blazar heating.   

In the simulations, we turn on blazar heating at redshift $z=5$ and adopt the redshift evolution of the blazar heating rate per co-moving volume $\dot{Q}(z)$ suggested by \citet{Chang2011} for $z<5$, i. e.  
\begin{align}
  \log_{10}&\left(\frac{\dot{Q}(z)}{\dot{Q}(z=0)}\right) = 0.0315\times[(1+z)^3 - 1] \nonumber \\ 
           & - 0.512\times[(1+z)^2 - 1] + 2.27\times[(1+z) - 1],
\label{eq:redshift_scaling}
\end{align}
where our assumed volumetric heating rates at the present epoch, $\dot{Q}(z=0)$, are summarised in Table \ref{tab:sim_models}. Our {\it weak} blazar heating model is slightly less efficient than the {\it standard} model in \citet{Chang2011}, while their {\it optimistic} model falls in between our {\it intermediate} and {\it strong} heating. For completeness, we also list the correction factors $\eta_{\rm sys}$ that correspond to our heating rates. $\eta_{\rm sys}$ \citep[defined in Eq.~16 of][]{Chang2011} is a parameter of order unity that corrects for remaining systematic uncertainties in the prediction of the heating rate due to the spectral variability of TeV blazars, their uncertain redshift evolution, and possible additional TeV sources that may contribute to the plasma instability heating.

Based on the rate given by Eq.~(\ref{eq:redshift_scaling}), the required heat injection for each SPH particle is implemented in the simulation code by accordingly updating the internal energy of all active SPH-particles at each time-step. The required estimate of the volume, $V=m_{\rm SPH}/\rho_{\rm SPH}$, occupied by a particle is computed as ratio of particle mass, $m_{\rm SPH}$, to SPH density estimate, $\rho_{\rm SPH}$.\footnote{In SPH simulations, the particle volumes obtained in this way typically do not add up exactly to the total volume of the simulation box. The sum of the volumes is about 5\% low in our runs, which potentially biases the heating rate low by the same fraction. We do not attempt to correct for this as this bias is small compared to the uncertainty in the blazar-heating rate.}

\begin{table}
\begin{tabular}{lccccc}
\hline
simulation model & $\dot{Q}(z=0)$ & $\eta_{\rm sys}$\\
& $[\,$eV Gyr$^{-1}$ cm$^{-3}]$ & \\
\hline
no blazar heating & 0 & 0\\
weak blazar heating & $5.8 \times 10^{-8}$ & 0.66\\
intermediate blazar heating & $1.08 \times 10^{-7}$ & 1.23\\
strong blazar heating & $1.62 \times 10^{-7}$ & 1.85\\
\hline
\end{tabular}
\caption{Assumed volumetric blazar heating rates at $z=0$ in our different simulations. The redshift evolution is in all cases given by Eq. (\ref{eq:redshift_scaling}). Also listed are the values of  $\eta_{\rm sys}$ corresponding to our heating rates.}
\label{tab:sim_models}
\end{table}

\subsection{Simulating the Lyman-$\alpha$ forest}
\label{sec:ly_alpha_methods}

\begin{figure*}
\centerline{\includegraphics[width=\linewidth]{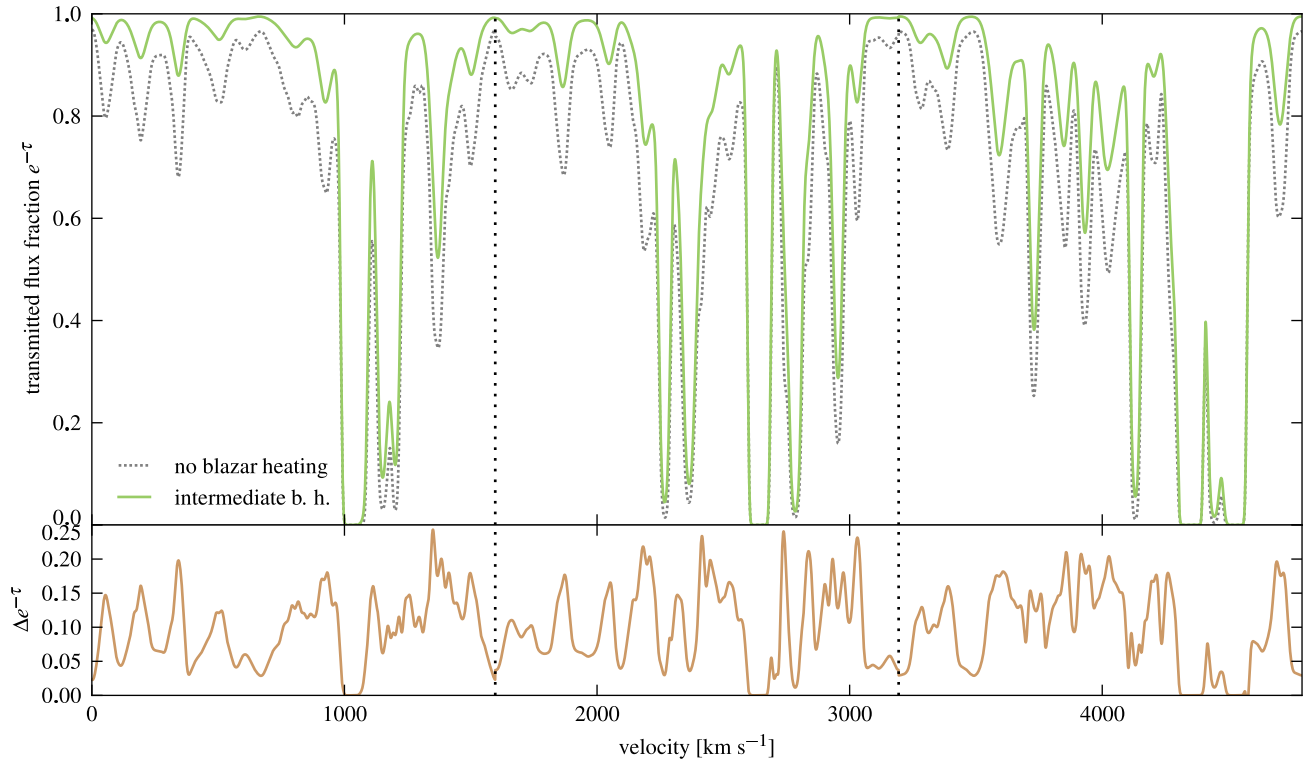}}
\caption{The Lyman-$\alpha$ forest in simulations with {\it intermediate} and without blazar heating at redshift $z=3$. More precisely, the {\it top panel} shows the transmitted flux fraction as a function of velocity shift for three randomly selected lines-of-sight through the simulation box. The three lines-of-sight were stitched together at suitable positions (indicated by the {\it vertical dotted} lines) to allow a better comparison to real spectra. Clearly, Lyman-$\alpha$ absorption lines are significantly deeper in the run without blazar heating. The {\it bottom panel} shows the increase of the transmitted flux fraction due to blazars. The spectra were computed based on the FG'09 UV background.}
\label{fig:lya_spec}
\end{figure*}

In order to be able to compare the IGM predicted by our simulations to observations of the Lyman-$\alpha$ forest, we need to compute synthetic Lyman-$\alpha$ forest absorption spectra. We do this by selecting a large number of randomly placed lines of sight through individual outputs of the simulation box, along directions parallel to one of the coordinate axes (randomly selected among $x$, $y$, and $z$), and each represented by 2048 pixels. Next, we compute the neutral hydrogen density, temperature and velocity fields of the IGM along these lines of sight by adding up the density contributions and averaging the temperatures and velocities of all SPH particles whose smoothing lengths are intersected. The calculation of the spectra then accounts for Doppler-shifts due to bulk flows of the gas along the line of sight as well as for thermal broadening of the Lyman-$\alpha$ resonance in the rest frame of absorbing gas \citep[see e.g.][]{Bolton2007}. This procedure yields the optical depth, $\tau$, for Lyman-$\alpha$ absorption as a function of velocity offset along each line of sight. This can then be easily converted into normalised transmitted flux, $F=e^{-\tau}$, as a function of wavelength or redshift. Examples of simulated absorption spectra, obtained in this way, are shown in Fig.~\ref{fig:lya_spec}.

\section{Results}
\label{sec:results}

\subsection{The thermal history and the temperature-density relation of the intergalactic medium}
\label{sec:th_eos_igm}

\begin{figure}
\centerline{\includegraphics[width=\linewidth]{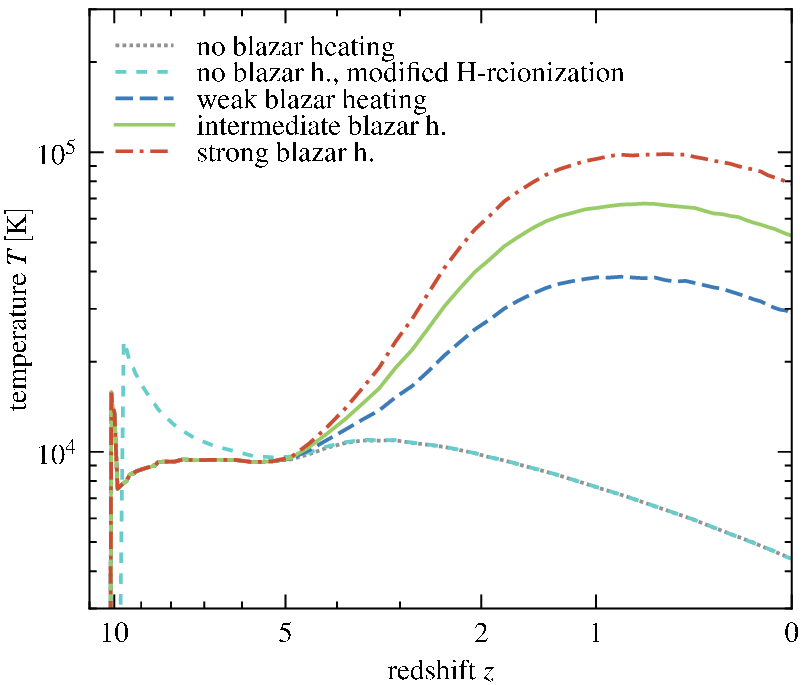}}
\caption{IGM temperatures in simulations with and without blazar heating as a function of redshift $z$. Shown are the median temperatures of the gas particles with densities between 0.95 and 1.05 times the mean cosmic baryon density. At low redshift, blazar heating strongly boosts IGM temperatures. For simulations without blazar heating, the temperature evolutions are shown for two different methods of treating hydrogen reionisation. We show that the IGM temperature at redshifts $z<5$ does not depend on the implementation of hydrogen reionisation. The $x$-axis was chosen to be linear in $\log_{10}(1+z)$. The {\it no blazar heating} curve is somewhat difficult to identify since it is identical with the {\it blazar heating} curves for $z>5$ and almost identical with the {\it no  blazar heating, modified H-reionization} curve for $z<5$.}
\label{fig:temp_evo}
\end{figure}

\begin{figure*}
\centerline{\includegraphics[width=\linewidth]{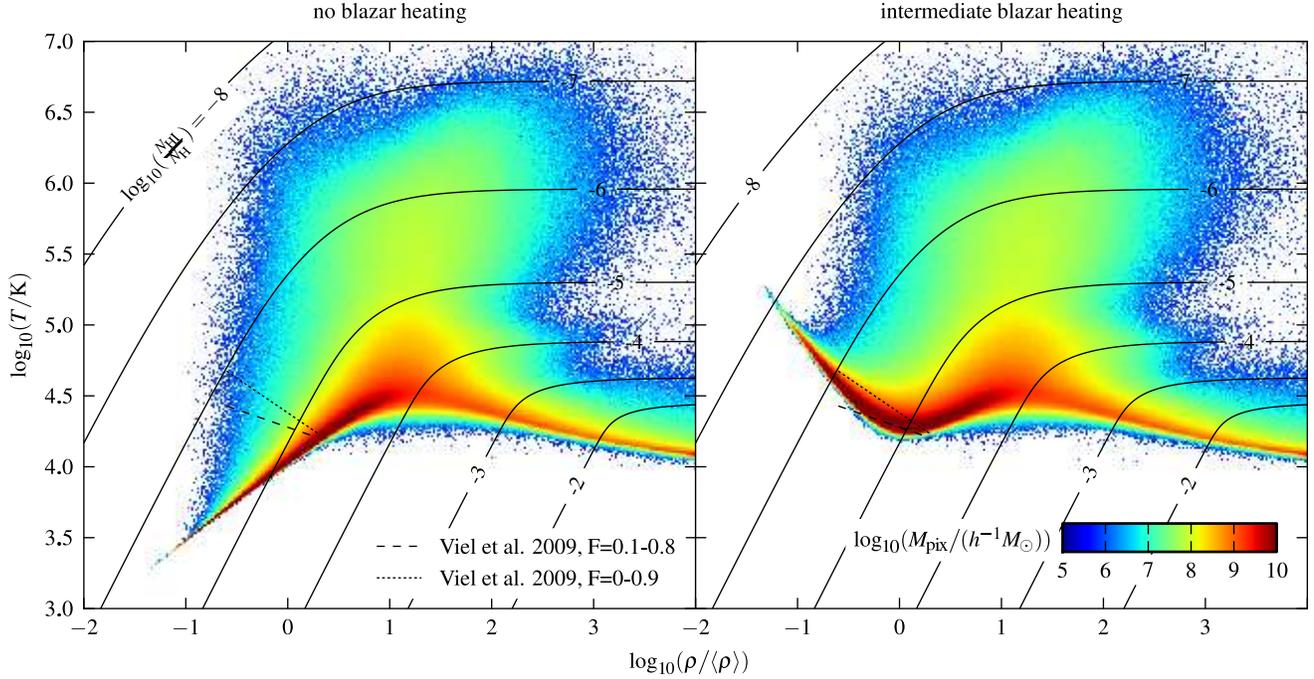}}
\caption{The gas density-temperature distribution in simulations without blazar heating ({\it left panel}) and with {\it intermediate} blazar heating ({\it right panel}) at redshift $z=3$. The gas mass per pixel $M_{\rm pix}$ is colour-coded. The colour scale spans 5 orders of magnitude. The contours indicate the neutral hydrogen fractions $\log_{10}(N_{\rm HI}/N_{\rm H})$ in the simulations. Also shown are the IGM T-$\rho$ relations obtained by \citet{Viel2009} based on an analysis of the Lyman-$\alpha$ forest transmitted flux distribution in two transmitted flux fraction ranges, $F=0.1-0.8$ and $F=0-0.9$, respectively, whose difference indicates the uncertainty in the data.}
\label{fig:rho_T}
\end{figure*}
\begin{figure*}
\centerline{\includegraphics[width=\linewidth]{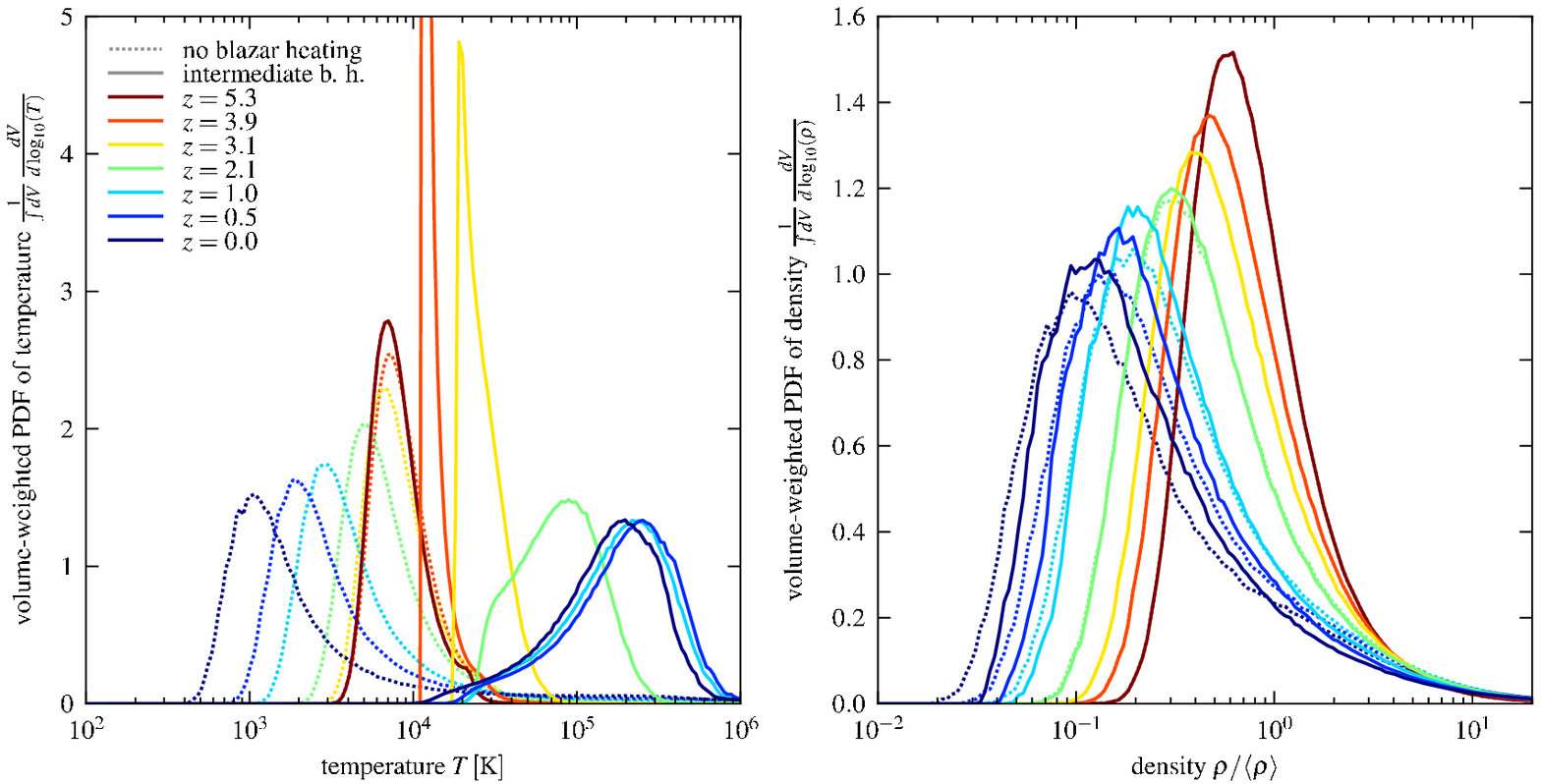}}
\caption{Volume-weighted probability distribution functions of temperature ({\it left panel}) and density ({\it right panel}) of the IGM. Results are shown for redshifts $z$ between 0 and 5.3, and for simulations without blazar heating ({\it dotted}) and with {\it intermediate} blazar heating ({\it solid}). In the latter run, the temperature distribution is shifted towards larger values once blazar heating becomes efficient for $z<5$. Without blazar heating, the temperature distribution evolves towards lower values once adiabatic cooling by the Hubble expansion exceeds heating by He {\sc ii}-reionisation at $z\lesssim 3.5$. With blazar heating, Hubble cooling starts to dominate only at $z\lesssim 0.5$. In both runs, the density distribution shifts towards lower values (in units of the mean baryon density $\langle \rho \rangle$) as a consequence of structure formation. It can also be clearly seen that blazar heating has a noticeable effect on the density distribution for $z\lesssim 2$.}
\label{fig:T_rho_pdfs}
\end{figure*}

In most of the models for the thermal history of the IGM found in the literature presently, hydrogen and helium reionisation injects the majority of the energy. Recent observations of the cosmic microwave background suggest that hydrogen was reionised around redshift $z_{\rm reion} \approx 10$ \citep{Komatsu2011}.\footnote{Their constraint $z_{\rm reion}=10.6\pm1.2$ is, however, weakened if a gradual reionisation or multiple epochs of H reionisation are considered.} The UV background adopted in our simulations is consistent with this finding and reionises hydrogen quickly at $z=10$. 

As already discussed in Sect. \ref{sec:simulations}, we have performed two runs with different treatments of H and He {\sc i} reionisation. They were designed to bracket the actual heat input during this epoch and were both performed without blazar heating. The thermal history of the IGM at the mean density is illustrated for both of these simulations in Fig.~\ref{fig:temp_evo}. Shortly after H and He {\sc i} reionisation, the temperatures are very different. However, adiabatic cooling due to the Hubble expansion, as well as by inverse-Compton scattering off  the CMB subsequently removes most of the injected energy. By $z=5$, this {\it loss-of-memory} effect has made the temperatures in both runs virtually identical, since we employ a model with sufficiently early H and He {\sc i} reionisation \citep{Hui2003}. Overall, this shows that the properties of the IGM at $z<5$ are comparatively insensitive to the treatment of H and He {\sc i} reionisation\footnote{Again, this statement may be weakened if a gradual reionisation or multiple epochs of H and He {\sc i} reionisation are assumed.}, essentially because this happens so early, whereas heat input at later times, e.g., due to He {\sc ii} reionisation, is more important in comparison. In the following, we only use simulations in which the $\Delta T_{\rm H,He I}$-boost was applied slightly before $z=10$.

For $z<5$, blazar heating becomes the dominant heating mechanism around and below mean densities in all simulations that include this effect. As a consequence, the IGM temperatures rise significantly in these runs and reach values that are about an order of magnitude larger than the temperature in the reference run without blazar heating. This dramatic effect is also illustrated in Fig.~\ref{fig:temp_evo}.

As suggested in \citet{Chang2011}, blazar heating is implemented in our simulations using a volumetric heating rate that is independent of density. Consequently, less dense regions receive more heating energy per unit gas mass, causing a larger temperature increase there. This results in an inverted $T-\rho$ relation of the IGM, as can be clearly seen in Fig.~\ref{fig:rho_T}, which shows the distribution of gas particles in the density and temperature plane at redshift $z=3$, both for the simulation with {\it intermediate} blazar heating and for the reference run without blazar heating. In the latter case, most of the low-density IGM follows a power-law T-$\rho$ relation, with a slope that is set by photoheating and adiabatic expansion and compression of the IGM due to structure formation. When including blazar heating, the T-$\rho$ relation of the low-density IGM no longer has a constant slope. Instead, we find an inverted T-$\rho$ relation in underdense regions, which turns over in an T-$\rho$ relation with a positive slope at higher densities where blazar heating becomes progressively  less efficient. At $z=3$, this turn-over occurs roughly at mean density. It shifts, however, to somewhat larger overdensities at later times. 

\citet{Viel2009} found that the transmitted flux distributions of the observed Lyman-$\alpha$ forest are, indeed, best fit by an inverted T-$\rho$ relation. Their best fits based on the ranges $F=0.1-0.8$ and $F=0-0.9$ for the transmitted flux fraction are also indicted in the figure. The comparison is shown only for the range of overdensities that contributes most to these transmitted flux intervals. More importantly, the inverted T-$\rho$ relations inferred by \citet{Viel2009} are clearly not found in the simulation without blazar heating, while the simulation with {\it intermediate} blazar heating is in good agreement with these results.

Also shown in Fig.~\ref{fig:rho_T} are contours of the neutral hydrogen fraction, which where computed under the assumption of ionisation equilibrium with the UV background. Obviously, the inverted T-$\rho$ relation reaches lower neutral hydrogen fractions for low densities. Hence, we expect significant changes in the Lyman-$\alpha$ forest. This will be explored in more detail in the following sections.

Overall, Fig.~\ref{fig:rho_T} shows that the distribution of IGM mass in temperature and density is significantly altered by blazar heating. The changes are, however, even more dramatic in the distribution of the cosmic volume in temperature and density. This is illustrated in Fig.~\ref{fig:T_rho_pdfs}. At redshift $z=5.3$, before blazar heating is turned on, both simulations naturally have the same temperature and density distributions. Subsequently, due to ongoing structure formation, most of the volume evolves towards lower density compared to the mean density. Such underdense regions evolve, however, very differently in runs with and without blazar heating. In the latter case, they cool adiabatically by the Hubble expansion once the heating by He {\sc ii} reionisation becomes subdominant at $z\lesssim 3.5$. In contrast, blazars heat these regions efficiently in the former case. Together, this results in fundamentally different evolutions of the temperature distributions in runs with and without blazar heating. By redshift $z=0$, the IGM that permeates most of the volume astonishingly is two orders of magnitude hotter when blazar heating is included. At low redshift, the differences in the density distribution also become noticeable. In particular, there is less volume above the mean density, which likely has implications for late-forming structures as suggested by \citet{Pfrommer2011}.

\begin{figure}
\centerline{\includegraphics[width=\linewidth]{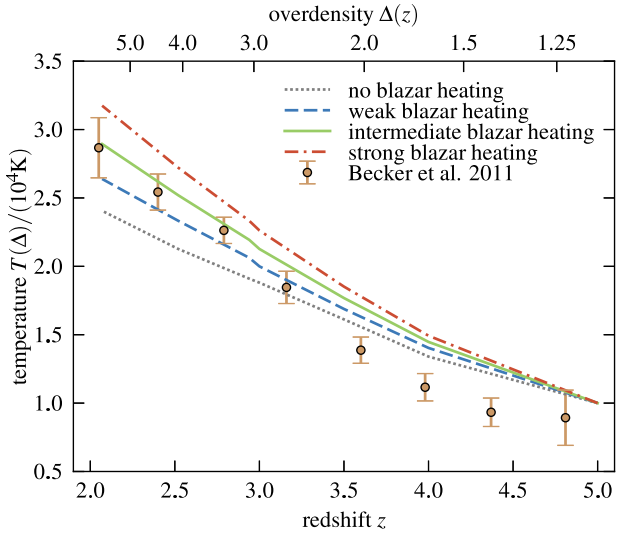}}
\caption{We compare IGM temperatures in simulations with and without blazar heating to constraints from Lyman-$\alpha$ forest observations by \citet{Becker2011}. The temperatures in the simulations were computed at the same redshift-dependent overdensity $\Delta(z)$ as in \citet{Becker2011}, so as to allow a direct comparison. At $z<3$, the additional heat input in the simulation with {\it intermediate} blazar-heating brings the IGM temperature into excellent agreement with the observations. At $z>3$, where the temperature increase due to blazar heating becomes progressively subdominant, the IGM is too hot in all our simulations, including the one without blazar heating. This discrepancy indicates tension between the He~{\sc ii} reionisation history implied by the FG'09 UV background and the Lyman-$\alpha$ forest temperature measurements.}
\label{fig:temp_evo_obs}
\end{figure}

In Fig.~\ref{fig:temp_evo_obs}, we directly compare the temperature in our simulations to temperature measurements based on Lyman-$\alpha$ forest observations \citep{Becker2011}. The observational constraints were obtained by relating the curvature of observed Lyman-$\alpha$ forest spectra to the IGM temperature at a specific redshift-dependent overdensity $\Delta(z)$. $\Delta(z)$ increases from $\Delta\sim1$ at $z=5$ to $\sim6$ at $z=2$. We computed the IGM temperatures in our simulation at the same overdensities to make them directly comparable. 

However, even at a fixed overdensity there is some ambiguity in the temperature definition. This can also be seen in Fig.~\ref{fig:rho_T}, which shows that the temperature distribution at a fixed overdensity between 3 and 6 is skewed and has a long tail towards high temperatures. Thus, the mean, the median, and the mode of the temperature distribution will differ. \citet{Becker2011} define the temperature $T(\Delta)$ by fitting a power law T-$\rho$ relation near the mean density and evaluate this fit at overdensity $\Delta$. In runs without blazar heating, the temperature obtained in this way closely tracks the mode of the mass-weighted temperature distribution at overdensity $\Delta$, which is the temperature where the distribution attains its maximum value. Thus, we use this temperature definition in Fig.~\ref{fig:temp_evo_obs}. 

For $z<3$, the simulation with {\it intermediate} blazar heating is in excellent agreement with the data, while the run without blazar heating has too low temperatures $T(\Delta)$. 

For $z>3$, where the temperature increase due to blazar heating becomes progressively unimportant, the IGM temperatures in all our simulations, including the one without blazar heating, exceed the \citet{Becker2011} values. This indicates tension between the early He {\sc ii} reionisation history implied by the FG'09 UV background model and the Lyman-$\alpha$ forest temperature measurements by \citet{Becker2011}. We would like to point out, however, that our blazar heating simulations are also in good agreement with temperature measurements based on the small-scale Lyman-$\alpha$ flux power spectrum \citep[e.g.,][]{Zaldarriaga2001,Lidz2009} in the range $2<z<3$, while our simulated temperatures are smaller than these measured values for $z > 3.5$. For $z > 3.5$, the simulated temperatures in all our runs, thus, fall in between constraints based on the flux curvuture statistics and the small-scale power spectrum, while being formaly inconsistent with both at the $1-\sigma$ level. This points to an unidentified systematics in either one or both of the methods for $z > 3.5$, where the absorbed fraction of the spectrum increases considerably, possibly suggesting an observational complication in measuring the temperature of the IGM at these redshifts.

\begin{figure}
\centerline{\includegraphics[width=\linewidth]{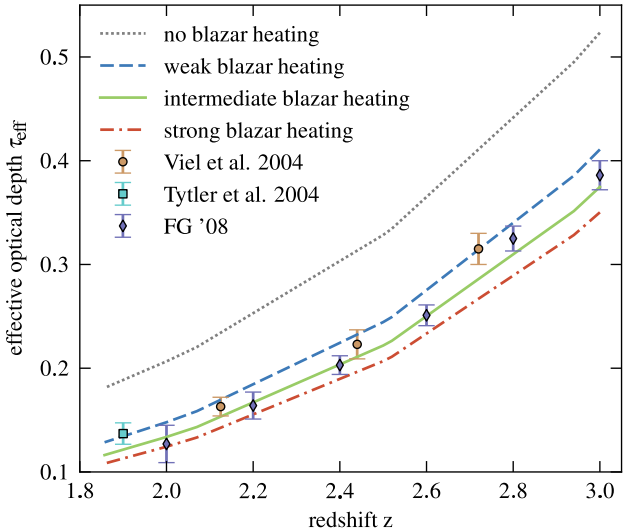}}
\caption{Effective optical depths, $\tau_{\rm eff}$, as a function of redshift for simulations with and without blazar heating. Observational data from FG'08, \citet{Viel2004}, and \citet{Tytler2004} are shown for reference. The simulations with blazar heating reproduce the observed optical depths very well. The results are based on the {\it self-consistent} UV background.}
\label{fig:tau_eff}
\end{figure}

Overall, the thermal histories of our simulations with and without blazar heating are remarkably different. In the next sections, we will explore how this affects predictions for the Lyman-$\alpha$ forest.

\subsection{Effective optical depths and photoionisation rates}
\label{sec:optical_depths}

To obtain detailed predictions for the Lyman-$\alpha$ forest in our simulations, we create synthetic spectra, as described in Sect.~\ref{sec:ly_alpha_methods}. Examples of such spectra are shown in Fig.~\ref{fig:lya_spec}. They are based on neutral hydrogen fractions computed under the assumption of ionisation equilibrium with the UV background which is described by the FG'09 model. In comparison to the simulation without blazar heating, there is significantly more flux transmitted in the runs that include blazars. Individual absorption lines are shallower and maximum values close to complete transmission are attained. Blazar heating boosts the transmitted flux most efficiently in lines with intermediate column densities, while fully absorbed lines are almost unaffected. Evidently, due to these substantial differences, the mean transmission in one of the simulations will be inconsistent with observational constraints. 

In the past, simulated absorption spectra for the Lyman-$\alpha$ forest were typically not taken at face value. Instead, the optical depths were rescaled to match the observed mean transmission values at every redshift separately. This can be justified if the actual UV background in the Universe is only poorly constrained. We checked that changing the UV background does not have a noticeable dynamical impact in the simulations. However, it will significantly change neutral hydrogen fractions, which are computed under the assumption of ionisation equilibrium in the presence of this background. In short, increasing the UV flux by some factor boosts the photoionisation rate by the same factor, and thereby lowers the lifetime of a freshly recombined hydrogen atom also by that factor. In equilibrium, the neutral hydrogen density (and thus also the optical depths) are proportional to the lifetime of a neutral hydrogen atom. Thus, by tuning the UV background one can rescale the optical depths. Commonly this was applied in the following way: using densities and internal energies from simulations, the corresponding optical depths were computed. These were then rescaled by an appropriate change of the UV background such that the resulting mean transmission was in agreement with observations. 

In the following, we check whether such a rescaling of the UV background is necessary to make our simulations match observed mean transmission values. Figure~\ref{fig:tau_eff} displays the effective optical depth, $\tau_{\rm eff}$, corresponding to the mean transmission defined by $e^{-\tau_{\rm eff}}=\langle e^{-\tau} \rangle$, as a function of redshift. Our simulations with blazar heating nicely reproduce the observed effective optical depths at all considered redshifts, while the simulation without blazar heating yields values that are significantly too large. This shows that blazar heating together with the recent FG'09 model of the cosmic UV background can fully account for the observed mean transmissions as a function of redshift. Instead, the run without blazar heating requires a rescaling of the UV background to get mean transmissions that are compatible with the data.

We perform our analysis of the simulated Lyman-$\alpha$ forest in the remainder of this paper in two independent ways, these are:

\begin {itemize}
 \item  We compute synthetic Lyman-$\alpha$ forest absorption spectra by self-consistently adopting the same FG'09 UV background model that was used to follow photoheating in the simulations. This UV background model agrees well with observational constraints from FG'08. In the following we refer to results obtained in this way to being based on the {\it self-consistent} UV background.\footnote{{\it Self-consistent} here refers to the fact that the same UV background model is used for the simulations and the Lyman-$\alpha$ forest analysis. The model is not based on following the UV sources in the simulations.} In this approach, all simulations are treated equally and the change in the Lyman-$\alpha$ forest properties due to blazar heating is not absorbed by artifically rescaling the mean transmission to the observed value. We will later show that this UV background allows us to simultaneously match observed IGM temperatures, Lyman-$\alpha$ forest line widths, and mean transmission values.

 \item In an alternative analysis, we artificially rescale the UV background at every redshift such that the optical depths of our simulated absorption spectra match the observed mean transmission. We, thus, compare the different simulations at the same mean transmission value. Results obtained in this way will be referred to being based on the {\it tuned} UV background. 
\end {itemize}   

\begin{figure*}
\centerline{\includegraphics[width=\linewidth]{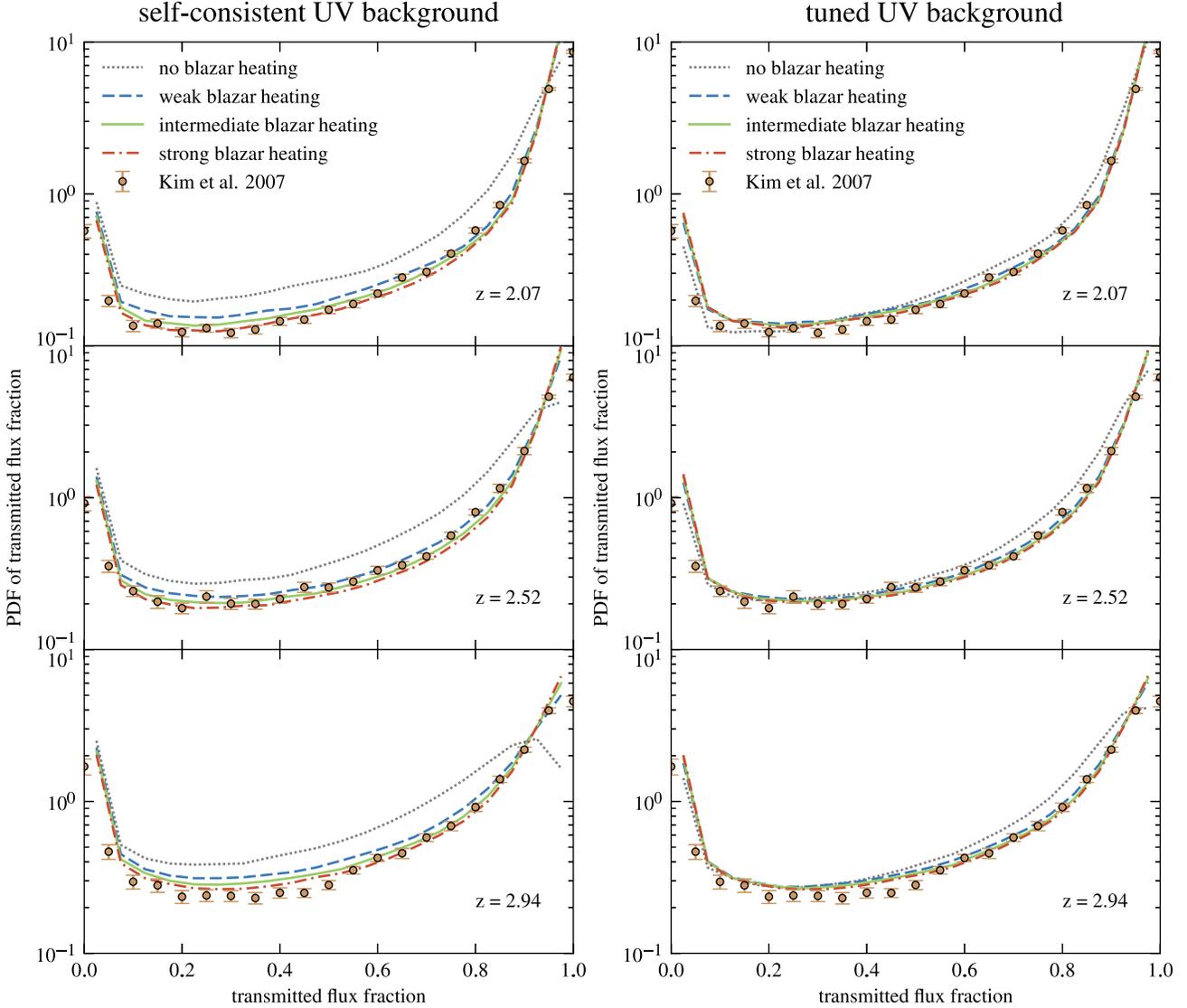}}
\caption{PDFs of the transmitted flux fraction for simulations with and without blazar heating. Results are shown for redshifts $z=2.07$, 2.52, and 2.94, and for the three different normalisations of the blazar heating rate. Observational constraints from \citet{Kim2007} are shown for reference. The {\it left} panels show results for the {\it self-consistent} FG'09 UV background, while the {\it right} panels show results for a UV background that, for each simulation and redshift, was {\it tuned} to match the observed mean transmission. In both cases, the simulations with blazar heating are in substantially better agreement with the data than the simulations without it.}
\label{fig:flux_pdfs}
\end{figure*}

\begin{figure*}
\centerline{\includegraphics[width=\linewidth]{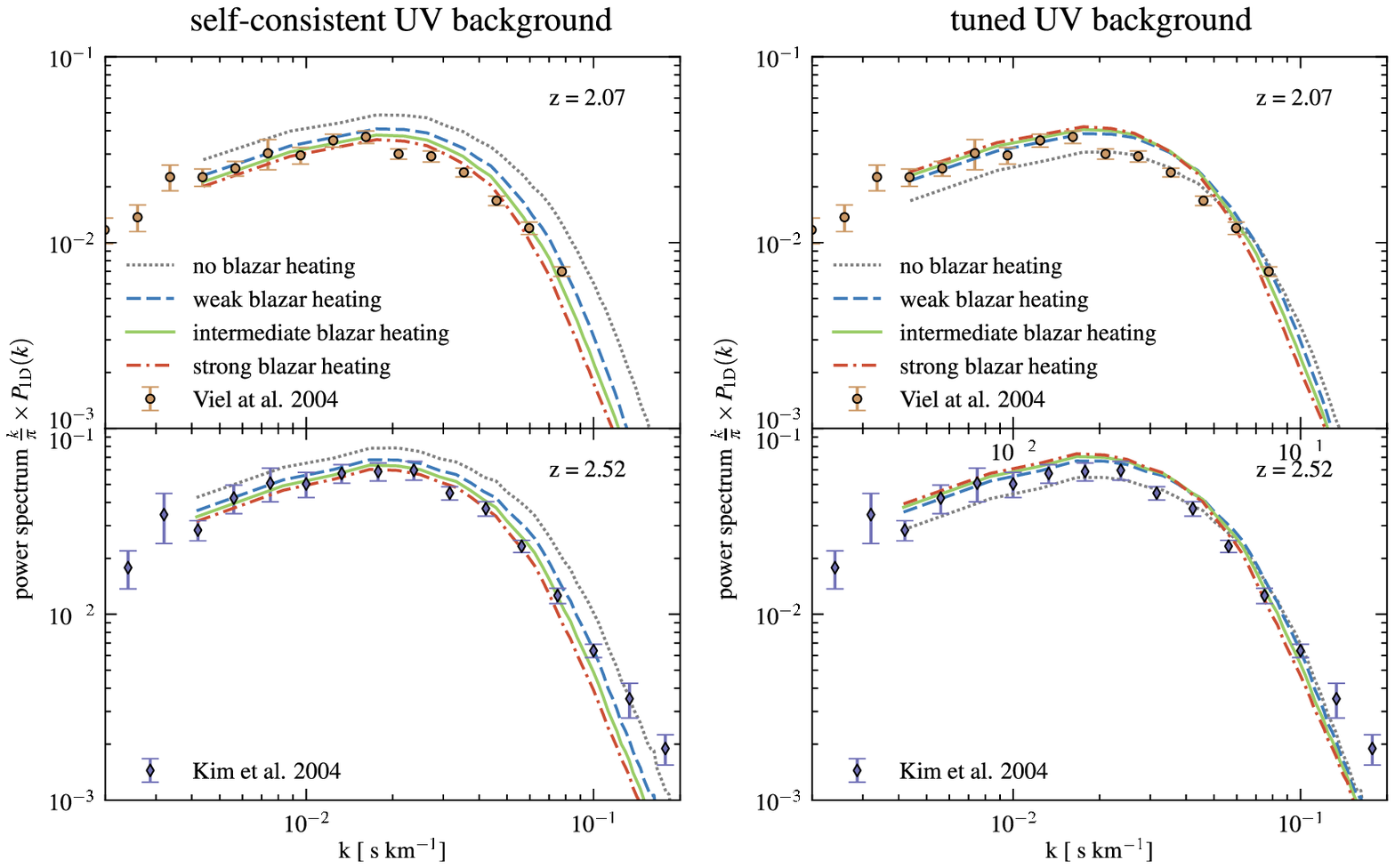}}
\caption{One-dimensional power spectra of transmitted flux contrast $\exp(-\tau)/\langle\exp(-\tau)\rangle-1$. Results are shown for simulations with and without blazar heating for redshifts $z=2.07$ and $2.52$. Observational constraints from \citet{Viel2004} and \citet{Kim2004erratum} for redshifts $z=2.125$ and 2.58, respectively, are shown for reference. The {\it left} panels show results for the {\it self-consistent} FG'09 UV background, while the {\it right} panels show results for a UV background that, for each simulation and redshift, was {\it tuned} to match the observed mean transmission. Including blazar heating in the simulations generally improves the agreement with this data, in particular for our {\it self-consistent} UV background analysis.}
\label{fig:power_spec}
\end{figure*}

Overall, these results show that in the run without blazar heating realisitic optical depths can only be obtained by rescaling the UV background. In the runs with blazar heating, the measured mean transmissions is, instead, {\em already} almost perfectly matched for all redshifts $1.9<z<3$ when using the UV background that was adopted in the simulations. Thus, results obtained by the {\it self-consistent} and the {\it tuned} analysis differ only very mildly for these runs.

\subsection{Transmitted flux probability distribution functions and power spectra}
\label{sec:flux_pdfs_power_spec}

We now move on to consider the PDFs and power spectra of the transmitted flux fraction at different redshifts. Both statistics are sensitive to the thermal state of the IGM \citep{Theuns2000}. Before analysing our synthetic spectra we smooth them to a resolution that is typical for real observations and add appropriate noise that accounts for Poisson distributed photon shot noise as well as a constant readout noise. We choose a Gaussian filter with $\sigma=3.5\,{\rm km \, s}^{-1}$ and assume a signal-to-noise ratio that corresponds to a value of 60 in $2.5\,{\rm km \, s}^{-1}$ pixels. In addition, readout noise is added with a standard deviation corresponding to $0.5\%$ of the continuum flux.

Figure~\ref{fig:flux_pdfs} displays transmitted flux PDFs for simulations with and without blazar heating and compares them to observations from \citet{Kim2007}. For our {\it self-consistent} UV background, we find that in the intermediate transmitted flux range $F=0.05-0.9$, blazar heating substantially lowers the PDF, so that it matches the data. Close to full transmission the behaviour is different. At $z=2.94$, the PDF from the simulation without blazars drops near $F=1$, which is not observed in the data. However, also the simulations with blazar heating deviate somewhat from the data at $F=1$. One should keep in mind though that the observed PDFs are very sensitive to uncertainties in the adopted continuum level there. An error of a few percent tends to broaden and lower sharp peaks in the PDF near $F=1$, like those found in the runs with blazar heating. This might account for a large part of the residual deviation between the data and our blazar runs. A discussion of the uncertainties in the PDF near $F=1$ can be found in \citet{Viel2009}.      

We now compare flux PDFs after rescaling all simulations to the observed mean transmission of the corresponding data sets by \citet{Kim2007}, i.e. we use the {\it tuned} UV background. As expected, we find that the differences between the different simulations are reduced. However, the runs with blazar heating are still in significantly better agreement with the observations. In particular, they do not overproduce the flux PDF in the $F=0.5-0.9$ transmitted flux range. \citet{Bolton2008} showed that lower values of the flux PDF in that range result from a flatter or inverted T-$\rho$ relation, which is naturally produced by blazar heating. In contrast, higher IGM temperatures alone are not sufficient to match observations there. Especially at $z=2.07$, part of the discrepancy in the run without blazar heating may, however, be due to finite box size effects \citep[see][]{Tytler2009}. At higher redshift, where the deviation is larger and finite box size effects are expected to be smaller, this is in contrast unlikely to significantly affect the comparison.

As a next step, we compute one-dimensional power spectra of the transmitted flux contrast, $\exp(-\tau)/\langle\exp(-\tau)\rangle-1$,
along our lines of sight. They are shown for runs with and without blazar heating at two different redshifts in Figure~\ref{fig:power_spec}, and are compared to observational constraints from \citet{Viel2004} and \citet{Kim2004erratum}. Based on the {\it self-consistent} UV background, we find that blazar heating drastically improves the agreement between simulated and measured values, reproducing both the shape and the normalisation of the observed power spectra nicely. The only remaining small deviation is in the $z=2.52$ simulation results at large $k$ values. At these small scales the measured power spectrum is, however, sensitive to residual contamination by metal lines that can result in excess power there \citep[e.g.,][]{McDonald2000}.

When comparing power spectra based on our tuned UV background analysis, we rescale optical depths to match the observed mean transmissions of the corresponding data sets by \citet{Viel2004} and \citet{Kim2004erratum}. We find that the runs with blazar heating are in significantly better agreement with the observations on large scales.\footnote{Finite box-size effects are not expected to singificantly affect our simulated power spectra on scales $k \gtrsim 0.005 \, \mathrm{s} \, \mathrm{km}^{-1}$ \citep[see, e.g.,][]{Tytler2009, Viel2004}.} On intermediate and small scales, where the difference caused by blazar heating is smaller for rescaled optical depths, all simulation models provide an acceptable match to the data.

To summarise, both the one and two-point statistics of the transmitted flux are in excellent agreement with observations once blazar heating is accounted for in simulation predictions, independent of whether or not we allow for a rescaling of the UV background to match the observed mean transmissions.

\subsection{Voigt profile fitting of Lyman-$\alpha$ lines}
\label{sec:line_fitting}

Another technique that has often been used to analyse observed Lyman-$\alpha$ forest spectra is to decompose them into individual absorption lines by fitting them with superpositions of Voigt profiles \citep[e.g.,][]{Hu1995,Kirkman1997}. Voigt profiles describe a thermally broadened spectral line of finite lifetime, in other words, they can be obtained by convolving a Lorentzian line profile with a Gaussian. The thermal broadening is usually quantified using the broadening parameter $b=(2kT/m_{\rm H})^{1/2}$, where $k$ is the Boltzmann constant, $T$ the temperature, and  $m_{\rm H}$ the mass of an hydrogen atom. Exploiting this relation, the distribution of line widths can be linked to the thermal state of the IGM. Thus, measuring the former at different redshifts allows tracing the thermal history of the IGM \citep{Schaye2000,Ricotti2000}.  

We applied this method to our synthetic Lyman-$\alpha$ forest spectra and compared the distributions of line-widths, $b$, and neutral hydrogen column densities, $N_{\rm HI}$, to observations. The Voigt-profile-decomposition was performed with the {\sc Autovp} code \citep{Dave1997}. Before applying it to our synthetic spectra we smoothed them and added noise as described in Sect.~\ref{sec:flux_pdfs_power_spec}. 

\begin{figure*}
\centerline{\includegraphics[width=\linewidth]{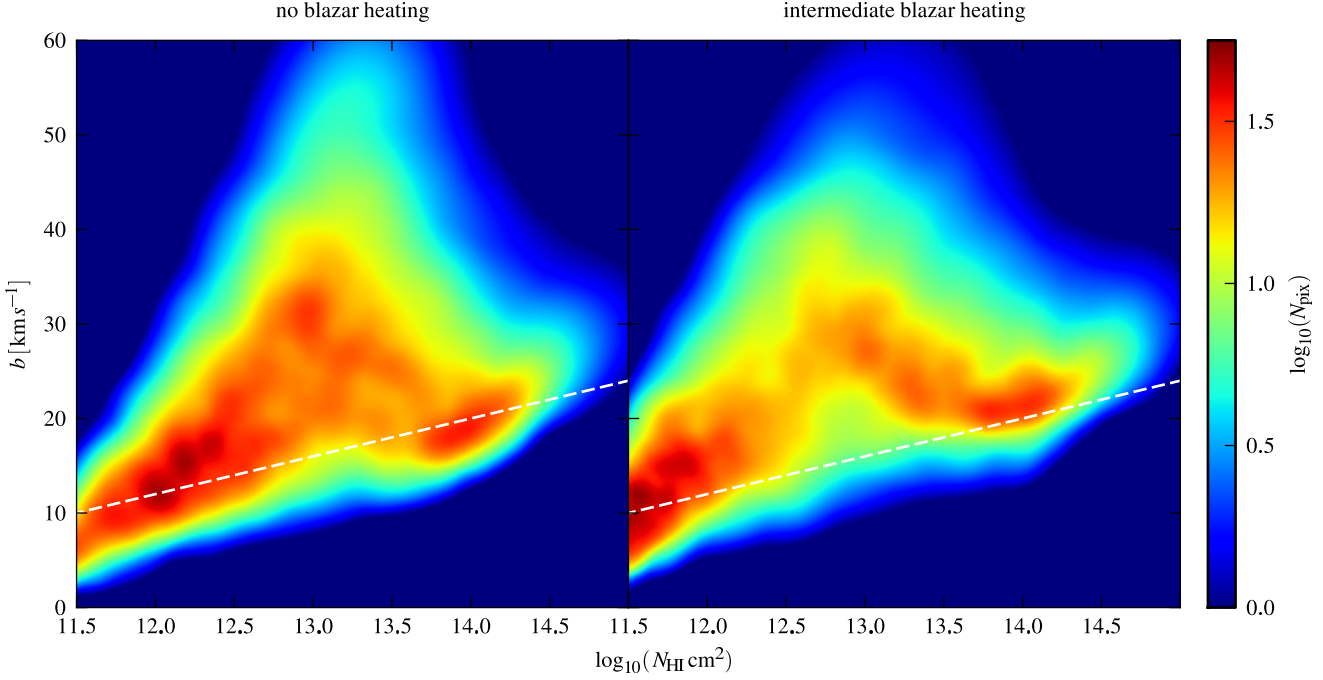}}
\caption{The distribution of Lyman-$\alpha$ absorption lines in column density, $N_{\rm HI}$, and line width, $b$, for simulations without ({\it left panel}) and with {\it intermediate} blazar heating ({\it right panel}) at redshift $z=3$. The frequency of lines per pixel (in arbitrary units) is colour-coded. Fits of roughly $10^4$ absorption lines in the synthetic spectra were used for these figures. Their distribution was adaptively smoothed using an SPH-like algorithm. The {\it white-dashed} lines indicate a fit to the observed lower $b$-envelope found by \citet{Kirkman1997} which is in good agreement with the simulation with {\it intermediate} blazar heating. The shown distributions are based on the {\it self-consistent} FG'09 UV background.}
\label{fig:b_nhi}
\end{figure*}

Figure~\ref{fig:b_nhi} shows the distribution of absorption lines in $b$ and $N_{\rm HI}$ for simulations without and with {\it intermediate} blazar heating. The lower $b$-cutoff, obtained from observations by \citet{Kirkman1997}, is shown for reference as a function of column density. As expected, the blazar heating boosts the IGM temperatures, which results in larger thermal broadening and accordingly shifts the lower $b$-cutoff to higher values. This brings the simulation prediction into excellent agreement with the measured cutoff if blazar heating is included. On the other hand, the run without blazar heating yields absorption lines with $b$-values that are too small. A direct comparison of the $b$-$N_{\rm HI}$ distribution of the observed sample and synthetic samples of similar size is given in Appendix \ref{sec:b_nhi_comp}. This comparison corroborates the superb agreement between the data and our {\it intermediate} blazar heating model.

\begin{figure*}
\centerline{\includegraphics[width=\linewidth]{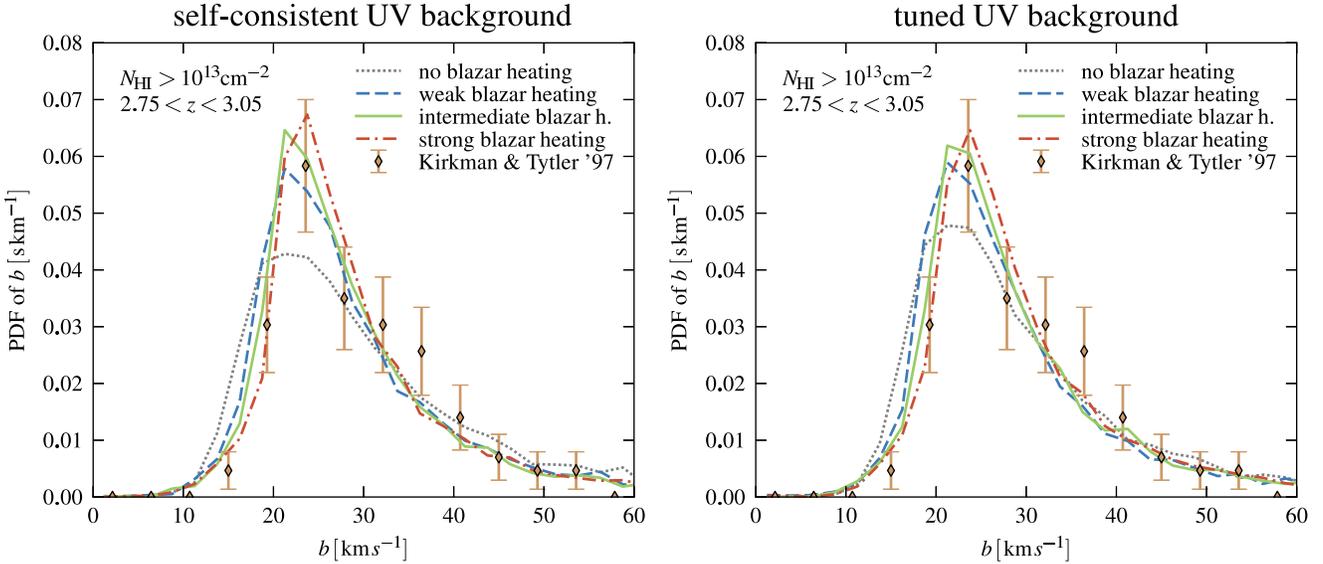}}
\caption{The normalised distribution function of Lyman-$\alpha$ line widths, $b$, for simulations with and without blazar heating at redshift $z=3$. Included are all lines with $N_{\rm HI}>10^{13} {\rm cm}^{-2}$. Observational constraints based on \citet{Kirkman1997} are shown for reference. They were obtained by using all lines from their Table 1 which were classified as Lyman-$\alpha$ lines, have $N_{\rm HI}>10^{13} {\rm cm}^{-2}$ and redshifts close to the simulation output ($2.75<z<3.05$). The {\it left} panel shows results for the {\it self-consistent} FG'09 UV background, while the {\it right} panel shows results for a UV background that was {\it tuned} for each simulation to match the observed mean transmission. The error bars assume Poissonian errors.}
\label{fig:b_pdf}
\end{figure*}

The shift of the lower $b$-cutoff can also be directly seen in the normalised distribution function of $b$-values, which is given in Figure~\ref{fig:b_pdf} for lines with $N_{\rm HI}>10^{13} {\rm cm}^{-2}$. Observational constraints based on \citet{Kirkman1997} are shown for reference. For both the {\it self-consistent} and the {\it tuned} UV background analysis, the data are only consistent with simulations that include blazar heating. Comparing the figure's panels also shows that this method relies only very weakly on the assumed photoionisation rate, as a rescaling of the optical depths only changes the column densities but not the widths of absorption lines. Thus, the only effect is a slight shift of the threshold value above which lines are included. Due to the weak dependence of the lower $b$-cutoff on column density, this does not strongly affect the results.

\begin{figure*}
\centerline{\includegraphics[width=\linewidth]{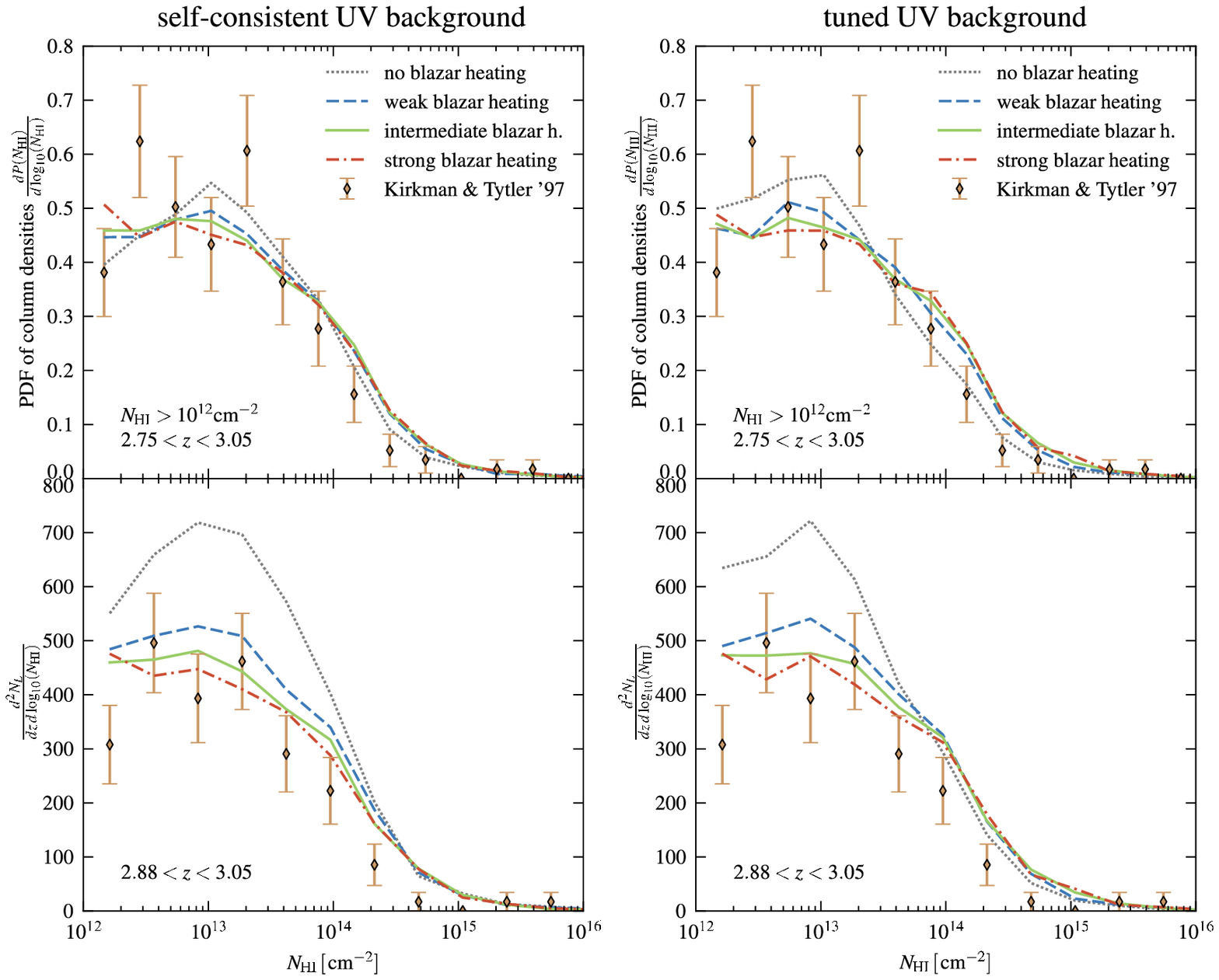}}
\caption{The normalised distribution of column densities $N_{\rm HI}$ of Lyman-$\alpha$ lines in simulations with and without blazar heating at redshift $z=3$ ({\it top panels}). Included are all lines with $N_{\rm HI}>10^{12} {\rm cm}^{-2}$. Observational constraints based on \citet{Kirkman1997} are given for reference. They were obtained by using all lines from their Table 1 which were classified as Lyman-$\alpha$ lines, have $N_{\rm HI}>10^{12} {\rm cm}^{-2}$ and redshifts close to the simulation output ($2.75<z<3.05$). The {\it lower panels} shows the number of lines per unit redshift and per dex in column density. Observational data from \citet{Kirkman1997} are shown for comparison, where only lines with $2.88<z<3.05$ were used to exclude gaps in their spectra at other redshifts. Results are shown for the {\it self-consistent} UV background ({\it left panels}) and the {\it tuned} UV background ({\it right panels}). All error bars assume Poissonian errors.}
\label{fig:nhi_pdf}
\end{figure*}

Figure~\ref{fig:nhi_pdf} compares both, the normalised distribution of column densities, as well as the frequency of lines with given column density per unit redshift to observations. The former distribution is only mildly affected by blazar heating\footnote{The distributions depend however significantly on the choice of the lower column density cutoff.}, and all simulations are in good agreement with the data. In contrast, the frequency of lines per unit redshift is much better reproduced in the runs with blazar heating. This is, of course, closely related to the better agreement of flux PDFs in these runs.  

In sum, we find that the simulation predictions match the observed properties of Lyman-$\alpha$ absorption systems impressively well when blazar heating is included. In particular, we would like to stress that the improved agreement with the observed lower cutoff of line-widths is a direct and independent confirmation of high IGM temperatures.

\section{Outlook and cosmological implications}
\label{sec:outlook}

In this section, we use our previous results on the effects of blazar heating and qualitatively discuss their impact on the local Lyman-$\alpha$ forest, on estimates of $\sigma_8$ (through the Lyman-$\alpha$ forest), on constraints of (sterile) neutrino masses, Helium {\sc ii} reionisation, formation of dwarfs and galaxy groups/clusters, and on particle acceleration at cosmological structure formation shocks. While we provide pertinent arguments for each point, we stress that each of these topics merits a thorough analyses to reliably quantify and confirm the mentioned effects, something which is beyond the scope of the present work.

\begin{figure*}
\centerline{\includegraphics[width=\linewidth]{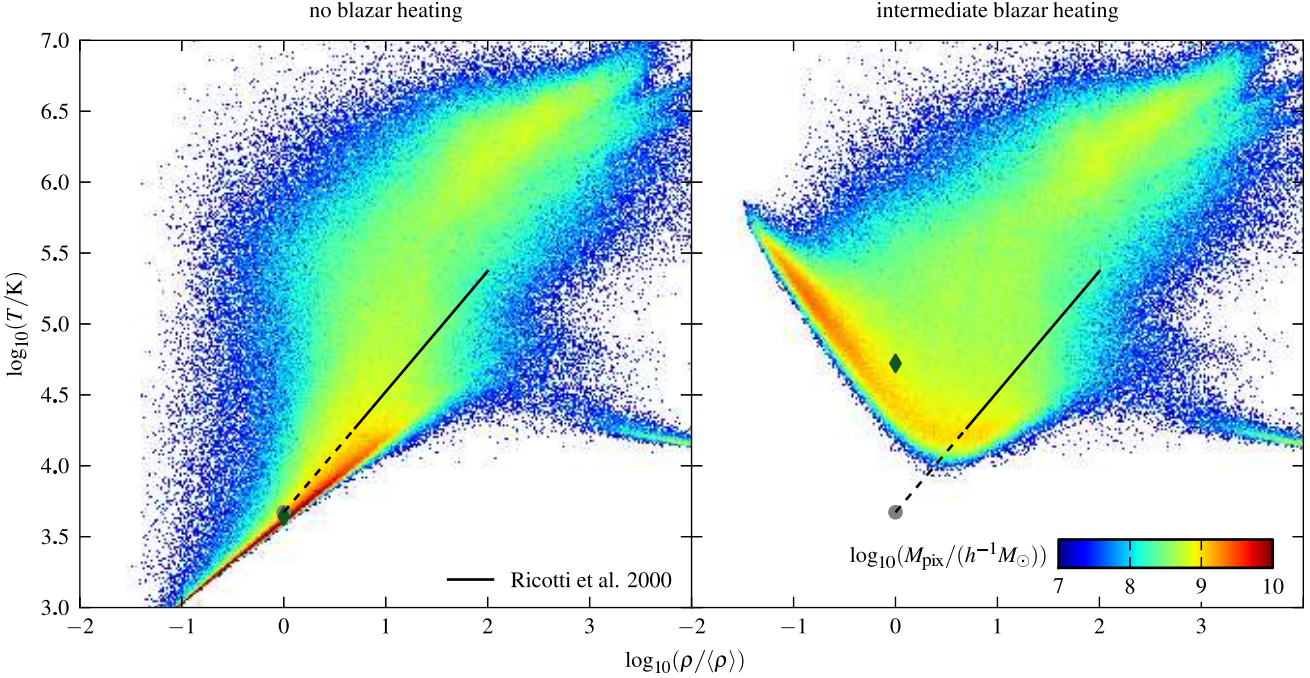}}
\caption{The gas density-temperature distribution in simulations without blazar heating ({\it left panel}) and with {\it intermediate} blazar heating ({\it right panel}) at redshift $z=0$. The gas mass per pixel $M_{\rm pix}$ is colour-coded. The colour scale spans 3 orders of magnitude. Also shown is the power-law T-$\rho$ relation obtained by \citet{Ricotti2000} from observations of the local Lyman-$\alpha$ forest that probes overdensities $5 \leq \Delta \leq 100$ (shown by the {\it solid} line). The {\it dashed} line indicates the extrapolation to the mean density ({\it grey circle}) assuming the same power-law relation holds there. The {\it green diamond} indicates the median gas temperatures at mean density, which is the same quantity that is shown in Fig.~\ref{fig:temp_evo} as a function of redshift. The thermal state of the IGM in the range where it is probed by the local Lyman-$\alpha$ forest is fairly similar in runs with and without blazar heating. However, while the power-law extrapolation to mean density works remarkably well without blazar heating, it completely fails in a blazar-heated universe in which the T-$\rho$ relation is tilted upwards at low density.}
\label{fig:rho_T_z0}
\end{figure*}

The trend that the Lyman-$\alpha$ forest probes larger overdensities at lower redshift, as mentioned in the discussion of Fig.~\ref{fig:temp_evo_obs}, continues down to $z=0$ \citep[see e.g.,][]{Schaye2001}. In particular, significant absorption lines in the local Lyman-$\alpha$ forest are expected to be due to absorbers at overdensities $5 \leq \Delta \leq 100$ \citep{Dave2010}, where the impact of blazar heating is strongly reduced. For example at an overdensity of 10 and $z=0$, we find only a $\sim 30\%$ temperature difference due to blazar heating instead of the factor of 10 difference that is shown in Fig.~\ref{fig:temp_evo} for mean density. We, thus, do not expect very drastic effects of blazar heating on the local Lyman-$\alpha$ forest. We would also like to point out that the latter does not yield direct constraints on the thermal state of the IGM at mean density. In particular, power-law extrapolations of the measured T-$\rho$ relation down to the mean density \citep[e.g][]{Ricotti2000} are invalid in a blazar-heated universe in which the T-$\rho$ relation is tilted. This is illustrated in Fig.~\ref{fig:rho_T_z0}. It may be interesting to explore in detail how blazar heating changes the statistics of the local Lyman-$\alpha$ forest in a future work.

One interesting aspect of our results is that by including blazar heating we were able to reproduce the observed transmitted flux power spectra well with simulations that adopt $\sigma_8=0.809$. This suggests that blazar heating could alleviate reported tensions between $\sigma_8$-values inferred  from Lyman-$\alpha$ forest studies and other cosmological constraints, most notably CMB observations \citep[e.g.,][]{Viel2006}. This tension is also present in the flux power spectra in our {\it tuned} UV background analysis (see {\it right} panels of Fig.~\ref{fig:power_spec}). There, the large-scale flux power in simulations without blazar heating falls short of the observed level. A joined analysis of flux PDFs of 27 high-resolution, high signal-to-noise ratio Lyman-$\alpha$ forest spectra \citep[the Large UVES Quasar Absorption Spectra, or LUQAS, sample by][]{Kim2004} and the SDSS flux power spectrum which is based on a large number of quasar spectra (3035) with low-resolution and low signal-to-noise \citep{McDonald2005} yields a larger value of $\sigma_8 = 0.86 \pm 0.03$ \citep{Viel2009} than an analysis of CMB data alone. This discrepancy appears to be related to the higher temperature $T_0$ that the flux power spectrum prefers and the fiducial T-$\rho$ relation slope of $\gamma\sim1$ around which the model was expanded. In fact, there is a degeneracy between the logarithmic slope of the T-$\rho$ relation at mean density, $\gamma=\rmn{d} \log P/\rmn{d}\log\rho$, and $\sigma_8$; larger values of $\gamma$ imply increasing values for $\sigma_8$ to be consistent with the data \citep[see Figure 2 in][]{Viel2009}. For a T-$\rho$ relation with a positive slope, $\gamma \gtrsim 1$, and fixing all other parameters, the inferred value of $\sigma_8\gtrsim 0.9$ would be in strong disagreement with values inferred from the CMB or other cosmological probes.  (We note, however, that such a procedure of fixing all but two parameters is only for illustrative purposes; in practice, all necessary parameters have to be allowed to vary simultaneously in order to identify the global best fit to the data.) Blazar heating provides a physically motivated scenario for an inverted T-$\rho$ relation with $\gamma\simeq0.5$, allowing the normalisation to match current concordance values for $\sigma_8\simeq0.8$ within 1$\sigma$ error bars.

Previous works that analysed Lyman-$\alpha$ power spectra in order to derive cosmological parameters and constraints on the mass of a sterile neutrino imposed Bayesian priors on the T-$\rho$ relation \citep{McDonald2005,Seljak2006}. Those were primarily motivated by theoretically expected limits, $0\lesssim \gamma-1\lesssim 0.6$ \citep{Hui1997}, rather than informed by Lyman-$\alpha$ forest observations themselves. These adopted priors in the parameter estimation algorithm excluded the possibility of an inverted T-$\rho$ relation during the marginalisation process. Hence, this may have lead to biased results if this shortcoming has not been absorbed somehow by the other (empirical) parameters within the minimisation procedure \citep[which employed a total of 34 parameters in][]{McDonald2005}. We have shown in Figure~\ref{fig:power_spec} that the cutoff scale of the one-dimensional power spectrum of the transmitted flux contrast is severely modified as a result of blazar heating and shifted towards lower values in comparison to models with photoheating alone. In principle, this could allow for a larger {\em intrinsic} cutoff in the transfer function as potentially imposed by a warm dark matter model and, hence, might weaken the constraints on the neutrino masses.

So far, He {\sc ii} reionisation seemed to be the only process able to substantially contribute to the late-time heating of the IGM. As a result, inhomogeneous He {\sc ii} reionisation has been studied to explain the high temperatures and the inverted T-$\rho$ relation of the IGM inferred from Lyman-$\alpha$ forest measurements \citep{Furlanetto08}. However, some of the properties that have been attributed to He {\sc ii} reionisation need to be re-evaluated within the framework of blazar heating, which can account for substantial heating for $z\lesssim3.5$. It will be an interesting challenge to constrain both, the characteristics of He {\sc ii} reionisation and blazar heating. If successful, this might even allow a determination of the redshift evolution of the blazar luminosity density by employing an analogous argument to that by \citet{Solt:82} for the thermal history of the IGM.  This could be achieved by combining the detailed temperature history as inferred from statistical analyses of the Lyman-$\alpha$ forest \citep[e.g.,][]{Becker2011} with the quasar luminosity density and improved constraints on the intrinsic quasar spectra above 4~Ry, the energy range that is of importance for He {\sc ii} reionisation, as a function of redshift.

A dramatically increasing minimum entropy  of the IGM for $z\lesssim3$ has a number of important implications for structure formation \citep[discussed in detail in][]{Pfrommer2011}. Such a redshift dependent entropy floor increases the characteristic halo mass below which dwarf galaxies cannot form by a factor of $\sim10$ at mean density over that found in models that include photoionisation alone (when adopting our successful {\em intermediate} blazar heating model which is only somewhat less efficient than the {\em optimistic} model studied in \citealt{Pfrommer2011}). This may help resolve the ``missing satellite problem'' in the Milky Way and the ``void phenomenon'' of the low observed abundances of dwarf satellites compared to cold dark matter simulations.

Simultaneously, the entropy injection by blazars is in some sense an amalgam of both the preheating and AGN feedback mechanisms for galaxy clusters. However, blazar heating operates on much larger distances than AGN feedback and provides a {\em time-dependent} entropy injection rate, peaking near $z\sim1$ (in contrast to the instantaneous heating previously assumed for preheating models). This suggests a scenario for the origin of the cool core/non-cool core bimodality in galaxy clusters and groups where early forming galaxy groups ($z\gtrsim1$) are unaffected due to the small level of entropy in the IGM at these times that can be quickly radiated away.  However, late forming groups do not have sufficient time to cool before the entropy is gravitationally reprocessed through successive mergers.

In Figure~\ref{fig:T_rho_pdfs}, we have shown the redshift evolution of the volume-weighted PDF of density in the blazar heating model, which starts to deviate from the model without blazar heating already at $z\sim2$. By $z\lesssim1$, the PDF is noticeably narrower in the blazar heating model, and the wings at low and high densities are suppressed. This suggests that blazar heating is indeed able to slow down non-linear structure formation on these scales. The degree to which this affects the luminosity function of dwarfs and the bimodality of central entropy values in galaxy clusters is subject of hydrodynamical cosmological simulations that we plan to carry out in forthcoming work.

Additionally, we have show in Fig.~\ref{fig:T_rho_pdfs}, that the median of the volume-weighted temperature PDF increases by more than a factor of 100 at $z=0$ as a result of blazar heating. This implies an increase of the sound speed, $c_\rmn{s}\propto \sqrt{kT}$, by a factor of $\sim 10$. Hence the strengths of the first shocks (as measured by the Mach number, $\mathcal{M} = v_\rmn{shock}/c_\rmn{s}$) which track the collapse of the IGM from voids onto large scale structure filaments and super-cluster regions are also reduced by a factor of $\sim 10$ \citep{Pfrommer+2006}. This should  have important implications for diffusive particle acceleration at structure formation shocks as the acceleration efficiency is a strong function of Mach number \citep[see e.g.,][and references therein]{Ensslin2007,Kang2010}. Similarly, this could also have interesting consequences for models of the resulting non-thermal emission from galaxy clusters, such as diffuse radio halos and relics.

\section{Summary and conclusions}
\label{sec:conclusions}

Recent studies discovered that TeV $\gamma$-ray emission from blazars can significantly heat the low-density IGM by annihilating and pair producing on the extragalactic background light.  The resulting high-energy electrons and positrons excite plasma beam instabilities which dissipate the pairs' energy locally \citep{Broderick2011,Chang2011}, leading to important implications for structure formation \citep{Pfrommer2011}. In this picture, the IGM acts as a cosmic heat bath that absorbs most of the bolometric energy of TeV blazars, which are ultimately powered by accretion onto their central super-massive black holes. Since the distributions of EBL photons and of TeV blazars are nearly homogeneous on cosmological scales, the resulting pair density is spatially homogeneous as well. As a result, the implied heating rate is independent of the IGM density.

We have included the expected heating term in state-of-the-art cosmological hydrodynamical simulations and investigated its impact on the thermal history of the IGM and the properties of the Lyman-$\alpha$ forest. We find that:

\begin{itemize}

\item In agreement with linear theory predictions by \citet{Chang2011}, the temperature of the low density IGM is strongly boosted by blazar heating, resulting in an inverted temperature-density relation as favoured by Lyman-$\alpha$ forest observations \citep{Viel2009}.

\item At $z<3$, when blazar heating is the dominant heating source, the IGM temperature, as a function of redshift, is also in good agreement with current constraints based on the curvature of observed Lyman-$\alpha$ forest spectra \citep{Becker2011}.

\item The mean transmissions predicted by runs with blazar heating are in excellent agreement with observations when employing the photoionisation rates predicted by up-to-date models of the cosmic UV background (FG'09). We adopt these rates in our {\em self-consistent} analysis which uses the same UV background for simulations and analysis. To allow for the uncertainties in the determination of the photoionisation rates, we additionally explore how results change if we {\em tune} the UV background to match the observed mean transmission values.

\item Blazar heating alters the distribution functions and power spectra of the transmitted flux and resolves discrepancies with observational constraints that exist when photoionisation is the only modelled source of heating. This improved match of simulations and data in our blazar heating models is also true if we rescale the photoionisation background to match the observed mean transmission at each redshift. This seems to be a unique property of the inverted $T$--$\rho$ relation \citep{Bolton2008} which is naturally produced by blazar heating.

\item Heating of the IGM by blazars shifts the lower cutoff of Lyman-$\alpha$ absorption line widths to larger values due to increased thermal broadening and decreases the abundance of weak lines with column densities $N_\rmn{HI}\lesssim3\times 10^{13} \,\rmn{cm}^{-2}$. This makes the lower cutoff and the abundance of lines per unit redshift and column densities consistent with observations and provides additional evidence for high IGM temperatures independent of whether or not we rescale to the observed mean transmission.

\item The inclusion of blazar heating brings the simulated large-scale flux power into agreement with the observed level. The simulation predictions are based on the most recent cosmological parameters which suggests that previous tensions in Lyman-$\alpha$ measurements of $\sigma_8$ when compared to other cosmological constraints are alleviated by this novel heating mechanism. Our results also support previous claims that the empirical assumption of an inverted temperature-density relation for the IGM already relieves the tension in the constraints on $\sigma_8$ \citep{Viel2009}.

\item Blazar heating dramatically alters the volume-weighted temperature PDF which implies that the strengths of first structure formation shocks during non-linear collapse are weakened. Future work is needed to work out the implications for the particle acceleration efficiencies and the resulting non-thermal emission from galaxy clusters. Modifications in the density PDF furthermore suggest that blazar heating may have interesting effects on structure formation, particularly on the scale of dwarf galaxies.

\end{itemize}

To summarise, including blazar heating in cosmological simulations results in properties of the IGM and the Lyman-$\alpha$ forest that are in excellent agreement with observations. In particular, we demonstrate in this work that our intermediate blazar heating model in combination with the FG'09 UV background represents the first successful, physically-motivated model that simultaneously reproduces all Lyman-$\alpha$ forest data (flux PDFs, power spectra, temperature evolution, line width and column density distributions) without the need to rescale the UV background at each redshift. This strengthens the case that this novel heating mechanism is indeed operating in our Universe, and at the same time it resolves long standing difficulties in our understanding of the Lyman-$\alpha$ forest.

This impressive improvement in modelling the Lyman-$\alpha$ forest appears to be a direct consequence of the peculiar properties of blazar heating. Its heating rate, which is independent of IGM density, naturally produces the inverted temperature-density relation that Lyman-$\alpha$ forest data suggest \citep{Viel2009}. The recent and continuous nature of the heating generates the redshift evolutions of the IGM temperature, the effective optical depth, the flux PDF, and flux power spectrum that accurately match observed trends for the redshift interval $2<z<3$ analysed here. Finally, the magnitude of the heating rate required by Lyman-$\alpha$ forest data approximately coincides with the total energy output of TeV blazars (or equivalently $\sim0.2\%$ of that of quasars).  While all these properties are quite unusual in the context of H and He {\sc i/ii} photoheating and traditional feedback processes, e.g., that of AGN or supernova driven winds, they are natural in the TeV-blazar heating mechanism. There is hence ample motivation for further studies of this scenario, which are also needed to establish in full detail how blazar heating affects cosmological constraints and structure formation over a range of scales.

\section*{Acknowledgements}
We would like to thank Claude-Andr\'{e} Faucher-Gigu\`{e}re and Mark Vogelsberger for making the photoionisation and photoheating rates that were used in our simulations available to us in numeric form. We are also grateful to Tom Abel, Joop Schaye, and Martin Haehnelt for helpful discussions. We thank our anonymous referee for thoughtful remarks that improved the manuscript. E.P. would like to acknowledge support by the DFG through Transregio 33. C.P. gratefully acknowledges financial support of the Klaus Tschira Foundation. A.E.B. and P.C. are supported by CITA. A.E.B. gratefully acknowledges the support of the Beatrice D. Tremaine Fellowship.


\appendix

\section{Comparing the $b-N_{\rm HI}$ distribution to observations}
\label{sec:b_nhi_comp}

In the following, we directly compare the simulated Lyman-$\alpha$ line widths, $b$, and neutral hydrogen column densities, $N_{\rm HI}$, to observations by \citet{Kirkman1997}. Rather than showing the distribution function as in Fig.~\ref{fig:b_nhi}, we present scatter plots of $b$ and $N_{\rm HI}$ values. All panels contain the same number of significant $N_{\rm HI}> 10^{12.5} {\rm cm}^{-2}$ lines as the observed sample. This facilitates a direct comparison. Note, however, that these results are more strongly affected by small-number statistics than those in Fig.~\ref{fig:b_nhi}. The figure displays results for simulations without blazar heating, results for simulations with {\it intermediate} blazar heating, as well as the observed sample.

Clearly, the lower $b$-cutoff is in much better agreement with the observations if blazar heating is included.  Interestingly, the region defined by $12.6 < \log_{10}(N_{\rm HI} \, {\rm cm}^2) < 13.1$ and $15 \, {\rm km \, s}^{-1}< b < 23 \, {\rm km \, s}^{-1}$, in which quite few lines fall in runs with blazar heating (due to the better statistics this can be more easily seen in the {\it right panel} of Fig.~\ref{fig:b_nhi}), is also very sparsely populated by the observed sample. This suggests that by including blazar heating also more subtle features of the $b-N_{\rm HI}$ distribution are better reproduced.

\begin{figure*}
\centerline{\includegraphics[width=\linewidth]{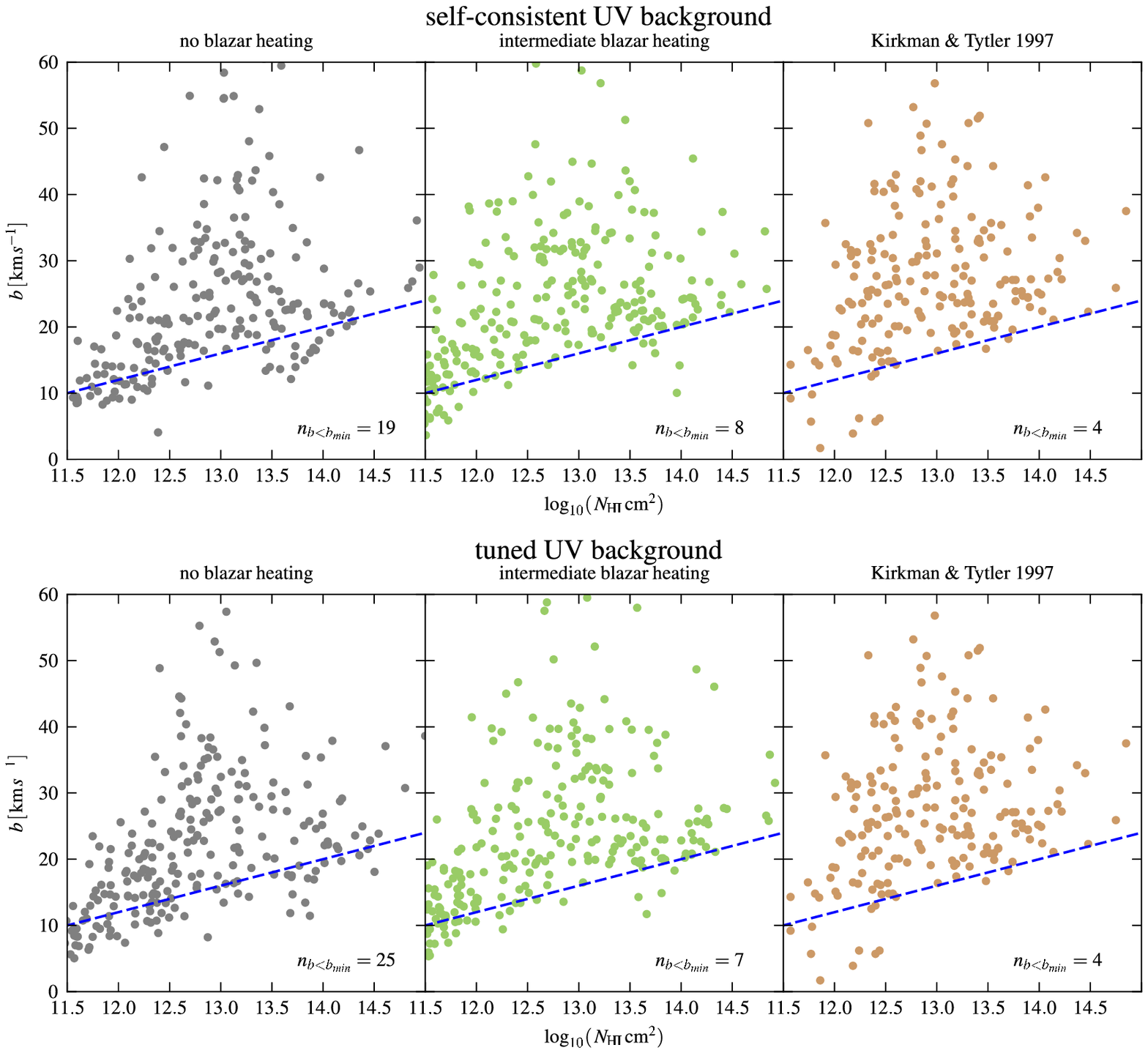}}
\caption{Scatter plot of the distribution of Lyman-$\alpha$ lines in column density, $N_{\rm HI}$, and line width, $b$, for simulations without ({\it left panels}) and with {\it intermediate} blazar heating ({\it middle panels}) at redshift $z=3$. In the {\it right panels}, data from \citet{Kirkman1997} are shown for comparison. All Lyman-$\alpha$ lines with redshifts $2.75<z<3.05$ listed in their Table 1 were included. The lengths of the (stitched) simulated lines-of-sight were chosen so as to yield the same number of significant ($N_{\rm HI}> 10^{12.5} {\rm cm}^{-2}$) Lyman-$\alpha$ absorption lines as in the observed sample. The {\it blue-dashed} lines indicate fits to the lower $b$-envelope found by \citet[given by their Eq.~4]{Kirkman1997}. The {\it upper} panels show results for the {\it self-consistent} FG'09 UV background, while the {\it lower} panels show results for a UV background that was {\it tuned} for each simulation to match the observed mean transmission. The number of lines $n_{b<b_{\rm min}}$ with $N_{\rm HI}> 10^{12.5} {\rm cm}^{-2}$ that fall below the fit to the lower $b$-envelope is indicated in each panel. For the observed sample one of these lines is not indicated in the figure as is has $N_{\rm HI}> 10^{15} {\rm cm}^{-2}$.}
\label{fig:b_nhi_scatter}
\end{figure*}

\bibliographystyle{mn2e}
\bibliography{master.bib}
\end{document}